\pdfoutput=1
\documentclass[PAPER, atlasdraft=false, coverpage=false, cernpreprint=true, UKenglish, texmf, orcidlogo]{atlasdoc}
\usepackage{atlaspackage}
\usepackage{atlasbiblatex}

\usepackage{atlasphysics}
\graphicspath{{logos/}{figures/}}

\usepackage{ANA-FTAG-2023-10-PAPER-defs}
\usepackage{multirow}
\usepackage{dirtytalk}
\usepackage{makecell}
\usepackage{siunitx}
\usepackage{adjustbox}

\usepackage{ANA-FTAG-2023-10-PAPER-defs}

\AtlasTitle{Measurement of the $b$-jet identification efficiency in dileptonic $t\bar{t}$ events using proton--proton collision data at $\sqrt{s}=13.6$~TeV collected with the ATLAS detector}

\AtlasAbstract{This paper presents the performance of the identification of jets containing $b$-hadrons ($b$-jets) for the GN2 algorithm, a transformer-based model for jet flavour tagging, using data collected by the ATLAS detector at the LHC. The analysis uses proton--proton collision data recorded in 2022 and 2023 at a centre-of-mass energy of $\sqrt{s} = 13.6$~TeV, corresponding to an integrated luminosity of 56~fb$^{-1}$. The $b$-jet identification efficiency and jet flavour composition are extracted simultaneously from a sample enriched in top-quark pair events ($t\bar{t}$). This efficiency is measured as a function of the jet transverse momentum in the range of 20--400~GeV and across six intervals of cumulative efficiency as derived in simulated $t\bar{t}$ events: [100\%, 90\%], [90\%, 85\%], [85\%, 77\%], [77\%, 70\%], [70\%, 65\%], and [65\%, 0\%]. The GN2 algorithm demonstrates significant performance improvements over its predecessor DL1d, achieving up to a factor of two (three) higher rejection of light-flavour (charm-flavour) jets at the same $b$-jet efficiency. The measured efficiencies in data are compared with simulation to derive correction factors ranging from 0.9 to 1.3. The total uncertainty is around 1\% for jets with transverse momentum larger than 60~GeV in the [65\%, 0\%] interval of cumulative efficiency.}
\AtlasRefCode{FTAG-2023-10}

\PreprintIdNumber{CERN-EP-2026-156}

\AtlasJournal{Eur.~Phys.~J.~C}
\AtlasCoverEgroupAnalysisTeam{atlas-ftag-2023-10-analysis-team@cern.ch}

\hypersetup{pdftitle={ATLAS document},pdfauthor={The ATLAS Collaboration}}

\begin{document}

\maketitle

\tableofcontents

\section{Introduction}
%
%

The ability to separate jets of different flavours, known as jet flavour tagging, is of crucial importance to the physics programme of the ATLAS experiment~\cite{PERF-2007-01} at the Large Hadron Collider (LHC)~\cite{Evans:2008zzb}. In particular, the ability to identify jets containing \textit{b}-hadrons (\bjets) or \textit{c}-hadrons (\cjets) and reject the large background of light-flavour jets plays a central role in the observation of the Higgs boson decay into bottom quarks and in the search for the Higgs boson decay into charm quarks~\cite{HIGG-2019-04, HIGG-2020-20, HIGG-2020-23}, many Standard Model (SM) precision measurements~\cite{STDM-2017-14, STDM-2018-43}, top-quark physics~\cite{TOPQ-2014-01, TOPQ-2024-12}, searches for Higgs boson pair production~\cite{HDBS-2021-18, HDBS-2022-02, HDBS-2019-27, HIGP-2025-10} and physics beyond the SM~\cite{EXOT-2021-15, HDBS-2021-03, EXOT-2017-02}.

Dedicated algorithms, referred to as jet flavour tagging algorithms, are exploited by the ATLAS Collaboration when analysing data to identify \bjets or \cjets. For the identification of \bjets, which is the focus of this paper, these algorithms typically rely on the long lifetime, high mass, large decay multiplicity of \textit{b}-hadrons, and the properties of the \bquark fragmentation. These physical characteristics are reflected by properties of charged-particle tracks, most notably large transverse and longitudinal impact parameters ($d_0$, $z_0$), which serve as the primary input for the tagging algorithms. The previous generation of algorithms utilised Deep Neural Networks~\cite{ATL-PHYS-PUB-2020-014} and Recurrent Neural Networks~\cite{FTAG-2019-07} to process these inputs. The current state-of-the-art algorithm used for Run~3 data analysis is GN2~\cite{FTAG-2023-05}, a General Network (GN) series machine-learning algorithm for flavour tagging based on the transformer architecture~\cite{transformers}.

GN2 predicts the probabilities of a jet being a \bjet ($p_b$), a \textit{c}-jet (containing a \textit{c}-hadron but no \textit{b}-hadron, $p_c$), a $\tau$-jet (containing a hadronically decaying $\tau$-lepton but no \textit{b}- or \textit{c}-hadrons, $p_\tau$), or a light-flavour jet ($p_u$) otherwise. A \bjet tagging discriminant $D_\texttt{GN2}$ is defined as:
\begin{equation*}
D_\texttt{GN2}= \ln \left( \frac{p_{b}}{f_{c} \cdot p_{c} + f_\tau \cdot p_\tau + (1-f_{c}-f_\tau) \cdot p_{u}}\right),
\end{equation*}
where the coefficients $f_c=0.2$ and $f_\tau=0.01$ were optimised to maximise the rejection of background jets in simulated \ttbar events. In addition to flavour prediction, GN2 is trained on auxiliary tasks, such as vertex prediction and track origin prediction~\cite{FTAG-2023-05}. These tasks improve the convergence of the training, enhance the overall flavour tagging performance, and increase the interpretability of the model. Evaluated in Monte Carlo (MC) simulation, GN2 provides a significant improvement on the previous algorithm (DL1d), with an increase in light-flavour jet rejection by a factor of two and \textit{c}-jet rejection by a factor of three at the same \bjet tagging efficiency.

This paper reports on the measurement of the GN2 \bjet tagging efficiency in data using proton--proton (\pp) collisions recorded at \sqs = 13.6 \TeV during 2022 and 2023. The measurement, also referred to as calibration, is used to derive correction factors, known as Scale Factors (SF), to account for differences in efficiency between data and MC simulation. The calibration is performed in Pseudo-Continuous \bjet Tagging (PCBT) bins, defined as exclusive ranges of the $D_\texttt{GN2}$ discriminant corresponding to intervals of cumulative \bjet tagging efficiency. A high-purity sample of \ttbar events is selected, where each top-quark decays as $t \to Wb$, with a branching ratio close to 100\%, and the $W$-boson decays leptonically, either in the electron channel or the muon channel ($W \to \ell \nu$). The selection requires exactly two isolated leptons with different flavours and opposite charges, along with two hadronic jets. This decay mode provides an abundant source of \bjets. Following the method in Ref.~\cite{FTAG-2018-01}, the \bjet tagging efficiency is extracted simultaneously with the jet flavour composition of the sample using a maximum likelihood approach. Systematic uncertainties are studied and incorporated into the final results.

This paper is structured as follows. Section~\ref{sec:detector} describes the ATLAS detector. Data and MC samples are discussed in Section~\ref{sec:DataMC}. The reconstruction of physics objects and event selection are covered in Section~\ref{sec:object}. Section~\ref{sec:fit} describes the calibration method, while Section~\ref{sec:uncertainties} discusses the uncertainties. Finally, results are shown in Section~\ref{sec:result}, followed by the conclusion in Section~\ref{sec:conclusion}.


%
\section{ATLAS detector}
\label{sec:detector}
%

%

The ATLAS detector~\cite{PERF-2007-01,GENR-2019-02} at the LHC covers nearly the entire solid angle around the collision point.\footnote{\AtlasCoordFootnote}
It consists of an inner tracking detector surrounded by a thin superconducting solenoid, electromagnetic and hadronic calorimeters,
and a muon spectrometer incorporating three large superconducting air-core toroidal magnets.

The inner-detector system (ID) is immersed in a \qty{2}{\tesla} axial magnetic field
and provides charged-particle tracking in the range \(|\eta| < 2.5\).
The high-granularity silicon pixel detector covers the interaction region and typically provides four measurements per track,
the first hit generally being in the insertable B-layer (IBL). %
It is followed by the SemiConductor Tracker (SCT), which usually provides eight measurements per track.
These silicon detectors are complemented by the transition radiation tracker (TRT),
which enables radially extended track reconstruction up to \(|\eta| = 2.0\).
The TRT also provides electron identification information
based on the fraction of hits (typically 30 in total) above a higher energy-deposit threshold corresponding to transition radiation.

The calorimeter system covers the pseudorapidity range \(|\eta| < 4.9\).
Within the region \(|\eta|< 3.2\), electromagnetic calorimetry is provided by barrel and
endcap high-granularity lead/liquid-argon (LAr) calorimeters,
with an additional thin LAr presampler covering \(|\eta| < 1.8\)
to correct for energy loss in material upstream of the calorimeters.
Hadronic calorimetry is provided by the steel/scintillator-tile calorimeter,
segmented into three barrel structures within \(|\eta| < 1.7\), and two copper/LAr hadronic endcap calorimeters.
The solid angle coverage is completed with forward copper/LAr and tungsten/LAr calorimeter modules
optimised for electromagnetic and hadronic energy measurements, respectively.

The muon spectrometer (MS) comprises separate trigger and
high-precision tracking chambers measuring the deflection of muons in a magnetic field generated by the superconducting air-core toroidal magnets.
The field integral of the toroids ranges between \num{2.0} and \qty{6.0}{\tesla\metre}
across most of the detector.
Three layers of precision chambers, each consisting of layers of monitored drift tubes, cover the region \(|\eta| < 2.7\),
except in the innermost layer of the endcap region, where layers of small-strip thin-gap chambers and Micromegas chambers both provide precision tracking in the region \(1.3 < |\eta| < 2.7\).
The muon trigger system covers the range \(|\eta| < 2.4\) with resistive-plate chambers in the barrel, thin-gap chambers in the endcap regions,
and the aforementioned small-strip thin-gap chambers and Micromegas chambers in the innermost layer of the endcap.

The luminosity is measured mainly by the LUCID--2 detector that records Cherenkov light produced in the quartz windows of photomultipliers located close to the beampipe.

Events were selected by the first-level trigger system implemented in custom hardware,
followed by selections made by algorithms implemented in software in the high-level trigger~\cite{TRIG-2022-01}.
The first-level trigger accepted events from the \qty{40}{\MHz} bunch crossings at a rate close to \qty{100}{\kHz},
which the high-level trigger further reduced in order to record complete events to disk at an average rate of about \qty{3}{\kHz}.

The Run-3 detector configuration benefits from several upgrades compared with that of Run~2 to maintain high detector performance at the higher pileup levels of Run~3.
The improvements include a new innermost layer of the muon spectrometer in the endcap region, which provides higher redundancy and a large reduction in fake muon triggers.
The trigger system also benefits from new digital electronics readout of the LAr calorimeters with significantly increased granularity.
Other updates and further details are provided in Ref.~\cite{GENR-2019-02}.

A software suite~\cite{SOFT-2022-02} is used in data simulation, in the reconstruction
and analysis of real and simulated data, in detector operations, and in the trigger and data acquisition
systems of the experiment.


%
\section{Data and simulated samples}
\label{sec:DataMC}
%

This measurement is performed using data collected during the 2022 and 2023 periods at $\sqrt{s} = 13.6$~\TeV, corresponding to an integrated luminosity of $56.0 \pm 1.1$~\ifb, after applying data quality requirements~\cite{ATL-DAPR-PUB-2023-001, ATL-DAPR-PUB-2024-001, DAPR-2018-01}. The luminosity is primarily measured by the LUCID-2 detector~\cite{LUCID2} located close to the beam pipe. Data are recorded using triggers that require the presence of a single electron (muon) with the minimum transverse momentum (\pt) exceeding a threshold of 26(24)~\GeV~\cite{TRIG-2022-01}.

MC simulated samples are used to estimate the yield of \ttbar production and other processes satisfying the selection criteria. Simulated \ttbar events are used to define the PCBT bins, while all MC samples are utilised in the extraction of the calibration SFs that correct for differences in \bjet tagging efficiency between simulation and data. The nominal \ttbar samples were generated with \POWHEGBOX~v2~\cite{Nason:2004rx, Frixione:2007vw, Alioli:2010xd} for the matrix element calculation at next-to-leading-order (NLO) in quantum chromodynamics (QCD) and interfaced to \PYTHIA[8.308]~\cite{Bierlich:2022pfr} with the A14 set of tuned parameters~\cite{ATL-PHYS-PUB-2014-021, ATL-PHYS-PUB-2012-003} for the parton showers, hadronisation and the underlying event. The \hdamp parameter, which regulates the high-\pt radiation in the matrix element to parton shower matching, is set to $1.5 \times m_{\textrm{top}}$. The \NNPDF[3.0nlo]~\cite{Ball:2014uwa} and \NNPDF[2.3lo]~\cite{Ball:2012cx} parton distribution functions (PDFs) are used in the matrix element calculation and the parton shower modelling, respectively. Three alternative \ttbar samples were produced for the evaluation of modelling uncertainties. The first one accounts for the modelling uncertainty in the parton shower radiation, in which the \hdamp parameter is set to $3.0 \times m_{\textrm{top}}$ and the \textsc{Var3} parameter variation group~\cite{ATL-PHYS-PUB-2016-020} is increased. The second set of alternative samples were generated with different choices for the $p_{\text{T,hard}}$ parameter~\cite{Hoche:2021mkv}, to evaluate the uncertainty in the matching between the NLO hard-scatter matrix element calculations and the parton shower~\cite{ATL-PHYS-PUB-2023-029}. The final alternative sample, which is used to evaluate the parton shower modelling uncertainties, was produced with \POWHEGBOX~v2 interfaced to \HERWIG[7.2.3]~\cite{Bellm:2019zci} with the default tune.

The non-\ttbar processes considered in the measurement either produce two prompt leptons or one prompt lepton and at least one non-prompt lepton candidate. These background processes include single top-quark production, single boson $V$+jets and diboson $VV$ production (where $V$ is either a \Zboson- or a \Wboson-boson). The nominal sample for the production of a single top-quark in association with a \Wboson boson ($Wt$) is simulated with \POWHEGBOX~v2 interfaced to \PYTHIA8.308 using the \say{diagram removal} scheme~\cite{Frixione:2008yi, Re:2010bp}. The samples for $s$- and $t$-channel production of a single top-quark were also produced with \POWHEGBOX~v2 interfaced to \PYTHIA8.308. Fully leptonically decaying diboson processes were simulated with \SHERPA~v2.2.14 and \SHERPA~v2.2.16~\cite{Gleisberg:2008ta, Hoeche:2009rj, Bothmann:2019yzt}. Samples of $Z$+jets, $W$+jets, and semileptonically decaying diboson processes, were produced with \SHERPA~v2.2.14. Alternative $Wt$ samples were produced with \POWHEGBOX~v2 interfaced to \PYTHIA8 using the \say{diagram subtraction} scheme~\cite{Frixione:2008yi, Re:2010bp} to assess the uncertainty related to the overlap between the \ttbar and $Wt$ production processes.

The effect of pile-up in the same and neighbouring bunch crossings was modelled in the simulation by overlaying the hard-scattering event with simulated inelastic \pp events~\cite{SIMU-2020-01} generated with~\EPOS~2.0.1.4~\cite{Porteboeuf:2010um,Pierog:2013ria} and~\PYTHIA~8.3. The events generated with~\EPOS used the~\EPOS LHC tune, while the~\PYTHIA events used the \NNPDF[2.3lo] set of PDFs and the A3 set of tuned parameters~\cite{ATL-PHYS-PUB-2016-017}.~\PYTHIA is used to simulate events that contain at least one high \pt jet, isolated photon or lepton from a $b$-hadron decay, while~\EPOS is used to simulate all remaining events in the sample. The events are generated independently and a filter is applied to the~\EPOS generated events to veto the event topologies simulated by~\PYTHIA. Events in the simulation were then reweighted to match the observed distribution of the average number of interactions per bunch crossing in data. The \EVTGEN~\cite{Lange:2001uf} package was used to model the decay of heavy flavour hadrons, except samples generated with \SHERPA that has its own model for hadron decays. All MC samples were processed through the full ATLAS detector simulation~\cite{SOFT-2010-01} based on \GEANT~\cite{Agostinelli:2002hh}, and are reconstructed using the same algorithms as in data.


%
%

%
\section{Object definition and event selection}
\label{sec:object}
%

Events are required to have at least one reconstructed collision vertex with two or more associated tracks with $\pt > 500~\MeV$. The primary vertex is defined as the vertex with the highest $\sum \pt^2$ of its associated tracks~\cite{ATL-PHYS-PUB-2019-015,IDTR-2021-01}. To obtain a \bjet-enriched sample, events compatible with originating from the dileptonic decay of a \ttbar pair are selected. This preselection requires exactly one electron candidate and one muon candidate with opposite charges and exactly two jets. At least one lepton must be identified in the trigger system and match one of the two offline lepton candidates. Corrections are applied to the simulation to account for known differences in trigger efficiency between the data and MC simulation.

Electron candidates are reconstructed from energy deposits in the Electromagnetic calorimeter matched to tracks in the ID~\cite{EGAM-2021-01}. They are required to satisfy $|\eta| < 2.47$, excluding the calorimeter transition region $1.37 < |\eta| < 1.52$, and have a transverse momentum $\pt > 28~\GeV$. Electrons must satisfy a tight identification criterion~\cite{EGAM-2021-01}. To select prompt electrons, candidates must fulfil tight isolation requirements~\cite{EGAM-2021-01}, which include the requirement to be isolated from nearby tracks in the ID and from additional calorimeter activity. Electron energies are calibrated using the \Zboson mass peak~\cite{EGAM-2021-02}. SFs are derived in $Z \rightarrow ee$ events and applied to the simulation to account for differences in reconstruction, identification, and isolation efficiencies between data and MC simulation.

Muon candidates are reconstructed from high-quality tracks in the MS matched to tracks in the ID~\cite{MUON-2018-03}. They are required to have $|\eta| < 2.5$ and $\pt > 28~\GeV$, and to satisfy a medium identification criterion~\cite{MUON-2018-03}. Similar to electron candidates, muon candidates must also fulfil tight isolation requirements. SFs are derived in $Z \rightarrow \mu\mu$ events and applied to the simulated events to account for differences in muon reconstruction, identification, and isolation efficiencies between data and MC simulation. Muon momenta are calibrated using the \Zboson mass peak~\cite{MUON-2022-01}.

Electron and muon candidates must originate from the primary vertex by requiring $|z_{0}| \sin \theta < 0.5~\text{mm}$ (where $\theta$ is the polar angle of the track), as well as $|d_{0}|/\sigma_{d_{0}} < 5$ for electrons (where $\sigma_{d_0}$ represents the uncertainty in the transverse impact parameter) or $|d_{0}|/\sigma_{d_{0}} < 3$ for muons.

Jets are reconstructed from particle-flow objects combining ID and calorimeter information~\cite{PERF-2015-09}, using the anti-$k_t$ algorithm~\cite{Cacciari:2008gp, Fastjet} with a radius parameter of $R = 0.4$. The jet four-momentum is calibrated to the particle level using a combination of simulation-based corrections and in situ calibrations~\cite{JETM-2018-05}. Jets are required to be central ($|\eta|<2.5$) and to have $\pt > 20~\GeV$. To suppress the contamination from pile-up jets, the jet vertex tagger (JVT) algorithm is applied to jets with $|\eta|<2.4$ and $\pt < 60~\GeV$~\cite{PERF-2014-03}, requiring a tight JVT selection criterion. In the MC simulation, reconstructed jets are labelled according to their flavour using truth information, following a procedure referred to as \textit{truth labelling}. A reconstructed jet is labelled as a \bjets if it contains at least one weakly decaying \textit{b}-hadron with $\pt > 5$~\GeV within a cone of size $\Delta R < 0.3$ around the jet axis~\cite{FTAG-2018-01}. If no \textit{b}-hadrons are found within the cone, matching to \textit{c}-hadrons is considered using the same selection criteria, followed by matching to $\tau$-leptons. All remaining jets are labelled as light-flavour jets. In the following, a reconstructed jet that has been truth labelled as a \bjet, \cjet or light-flavour jet is referred to as a \textit{truth} \bjet, \cjet or light-flavour jet, respectively.

Overlap removal is applied to avoid double-counting physics objects. To prevent double-counting electron energy deposits as jets, all reconstructed jets within $\Delta R < 0.2$ of a reconstructed electron are removed. If the nearest jet surviving the selection is within $\Delta R < 0.4$ of the electron, the electron is discarded. To minimise the fraction of muons from heavy-flavour decays inside jets mistakenly identified as prompt muons, muons are required to be separated by $\Delta R > 0.4$ from the nearest jet. In cases where a muon and a jet are reconstructed within $\Delta R < 0.4$ of one another, the muon is removed if the jet has at least three associated tracks; otherwise, the jet is removed. This ensures that high-energy muons undergoing significant energy loss in the calorimeter are not mistakenly considered as jets. %

The following preselection criteria are applied, on top of the trigger requirements, to select a sample enriched in \ttbar events:
\begin{itemize}
\item \textbf{Lepton selection}. Exactly two well-identified leptons of different flavour and opposite charges ($e\mu$) are required to suppress the background from $Z$+jets and non-prompt leptons events. To minimise the contamination from events where the two leptons originate from $\tau$-lepton decays, the electron--muon invariant mass $m_{e\mu}$ is required to be above $50~\GeV$.
\item \textbf{Jet selection}. Events with exactly two jets with $\pt > 20~\GeV$ are selected. This requirement suppresses the contribution of events with high-\pt light-flavour jets from final- or initial-state radiation or hadronic \Wboson-boson decays.
\item \textbf{Jet--lepton pair invariant mass selection}. There are two unique ways to pair the two leptons and the two jets. To best associate the objects originating from the same top-quark decay, the pairing that minimises $\sqrt{m^2_{j_1,\ell_i}+m^2_{j_2,\ell_j}}$ is chosen, where $j_1$ ($j_2$) is the leading(subleading) in $\pt$ of the jet, and $\ell_{i,j}$ are the two leptons. This pairing procedure selects the jet--lepton pairing most compatible with the top-quark decay topology, for which the invariant mass has a kinematic endpoint at around $m_{j,\ell} \lesssim 150~\GeV$. Both $m_{j_1,\ell}$ and $m_{j_2,\ell}$ are further required to be above $20~\GeV$ to avoid mis-modelling associated with very soft objects.
\end{itemize}
The mis-tagging rate of non-\bjet components in the MC simulation has been corrected with the procedure described in Section~\ref{sec:uncertainties}. After the correction, the data and MC comparison of the $\eta$ and \pt distributions of the leading and subleading jets for events satisfying the preselection are shown in Figure~\ref{fig:SELECTION-jet-kinematics-run3}. In the upper panel, the pre-fit MC distribution is shown as a breakdown of the \ttbar contribution and the non-\ttbar processes. In the ratio panel, the pre-fit MC distribution is compared with the post-fit MC distribution corrected by the measured \bjet tagging efficiency and jet flavour composition (measurement method described in Section~\ref{sec:fit}).

\begin{figure}
\centering
\subfloat[]{ \includegraphics[width=0.47\textwidth]{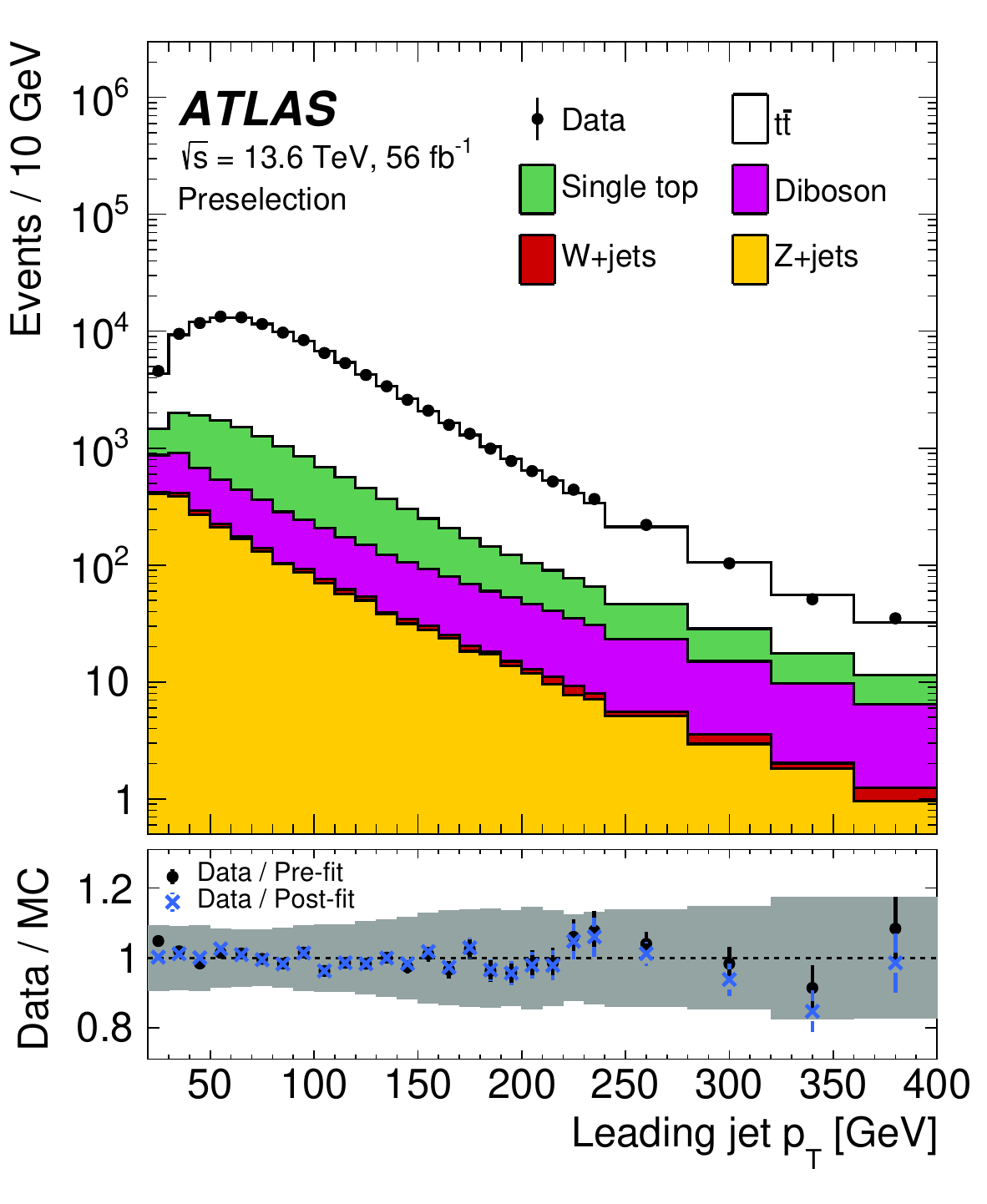}}
\subfloat[]{ \includegraphics[width=0.47\textwidth]{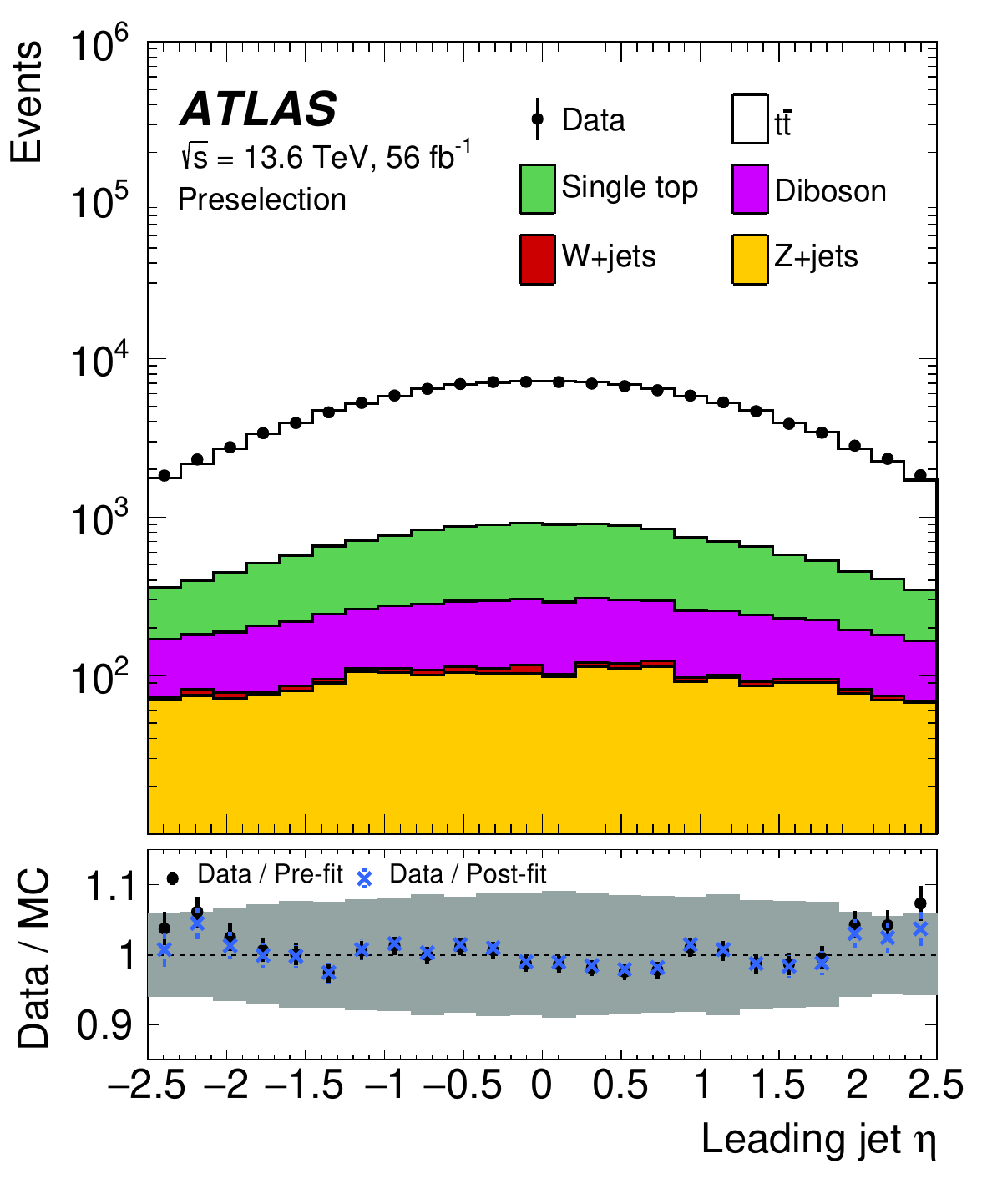}} \\
\subfloat[]{ \includegraphics[width=0.47\textwidth]{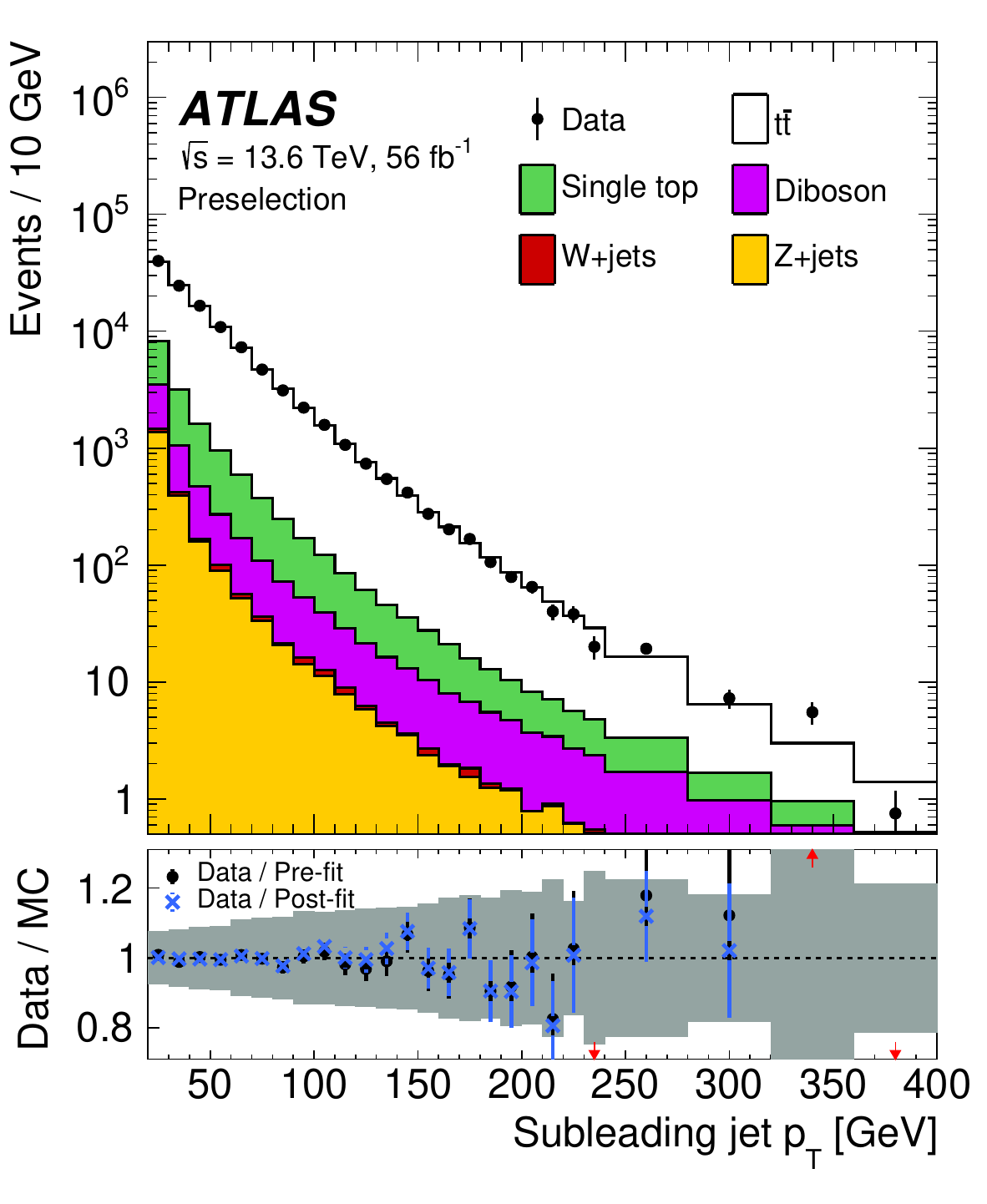}}
\subfloat[]{ \includegraphics[width=0.47\textwidth]{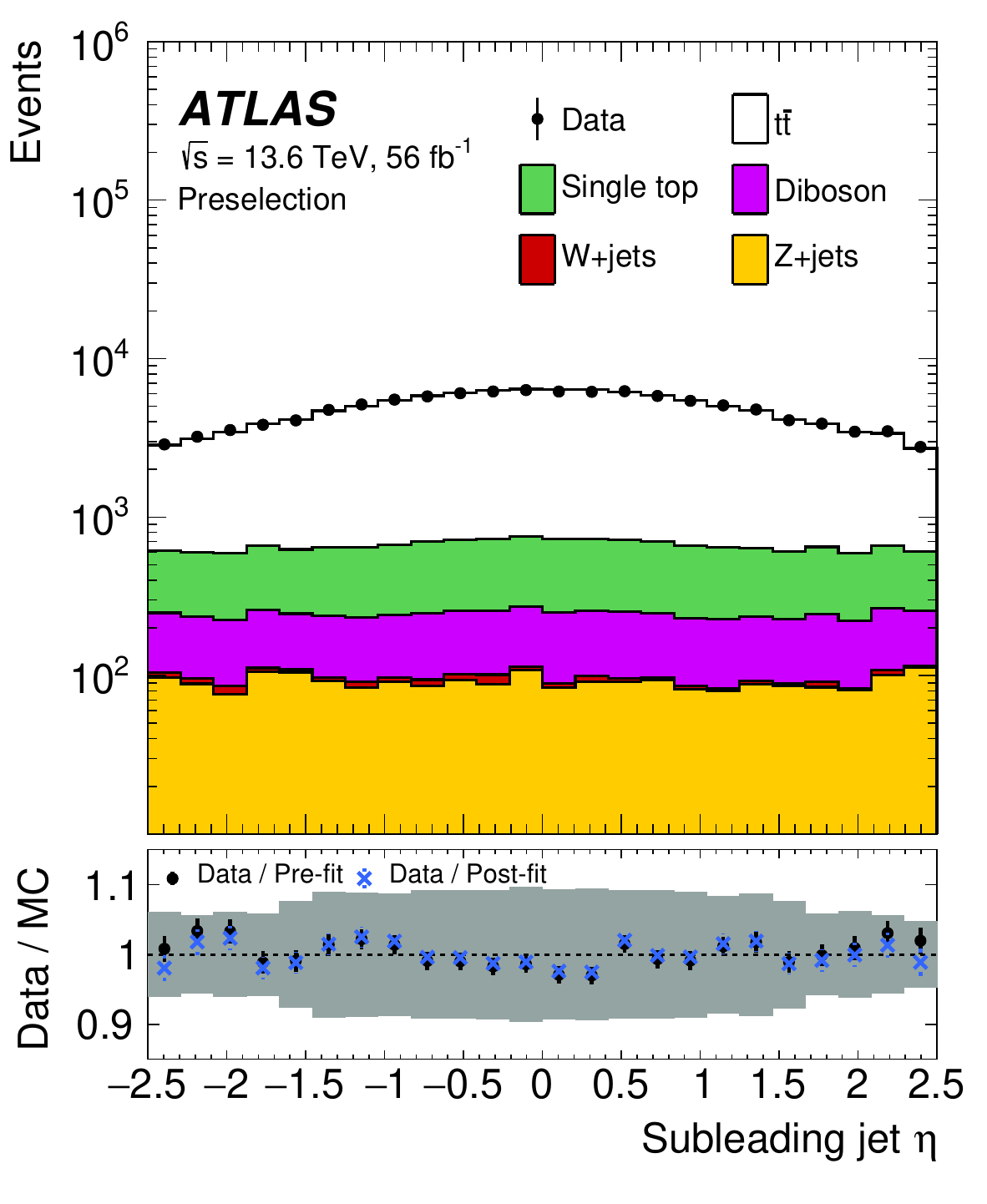}}
\caption{Distributions of the (a) leading jet \pT, (b) leading jet $\eta$, (c) subleading jet \pt and (d) subleading jet $\eta$, for events satisfying the selection criteria in data and MC samples. MC events are split into the \ttbar signal and the background processes. The total contribution of simulated MC events is normalised to the total data yield. The bottom panels show the data-to-MC ratio, where the black markers correspond to the pre-fit MC simulation and the blue markers correspond to the post-fit MC simulation. The error bars represent the statistical uncertainty of the data. The shaded band, displayed only in the bottom panel, is the sum in quadrature of MC statistical and systematic uncertainties (described in Section~\ref{sec:uncertainties}) in the pre-fit MC simulation.}
\label{fig:SELECTION-jet-kinematics-run3}
\end{figure}

Based on the truth flavour of the leading and subleading selected jets in simulation, events are labelled as $bb$, $bl$, $lb$, or $ll$. Here, $l$ refers to all non-$b$ jets, including \cjets, of which the contribution is one order of magnitude smaller than that of light-jets after the preselection. Specifically, $bb$ ($ll$) events are those where both (neither) of the selected jets are truth \bjets.

Events satisfying the preselection are further categorised depending on the jet--lepton invariant masses $m_{j_1,\ell}$ and $m_{j_2,\ell}$ shown in Figure~\ref{fig:SELECTION-mjl-distributions} to define a Signal Region (SR) enriched in $bb$ events and three Control Regions (CR) enriched in $lb$, $bl$, and $ll$ events. These control regions help to estimate the background contamination from mistagged events. The SR is defined by requiring both $m_{j_1,\ell}$ and $m_{j_2,\ell}$ to be below $175~\GeV$. In the SR, the $bb$ purity is approximately 58\%; the remaining 42\% consists of events where one or both jets are truth light or charm jets. The categorisation is summarised in Table~\ref{tab:SELECTION-analysis-region-definition} and the jet flavour compositions are shown in Figure~\ref{fig:SELECTION-true-flavour-fractions}.

\begin{figure}
\centering
\subfloat[]{ \includegraphics[width=0.47\textwidth]{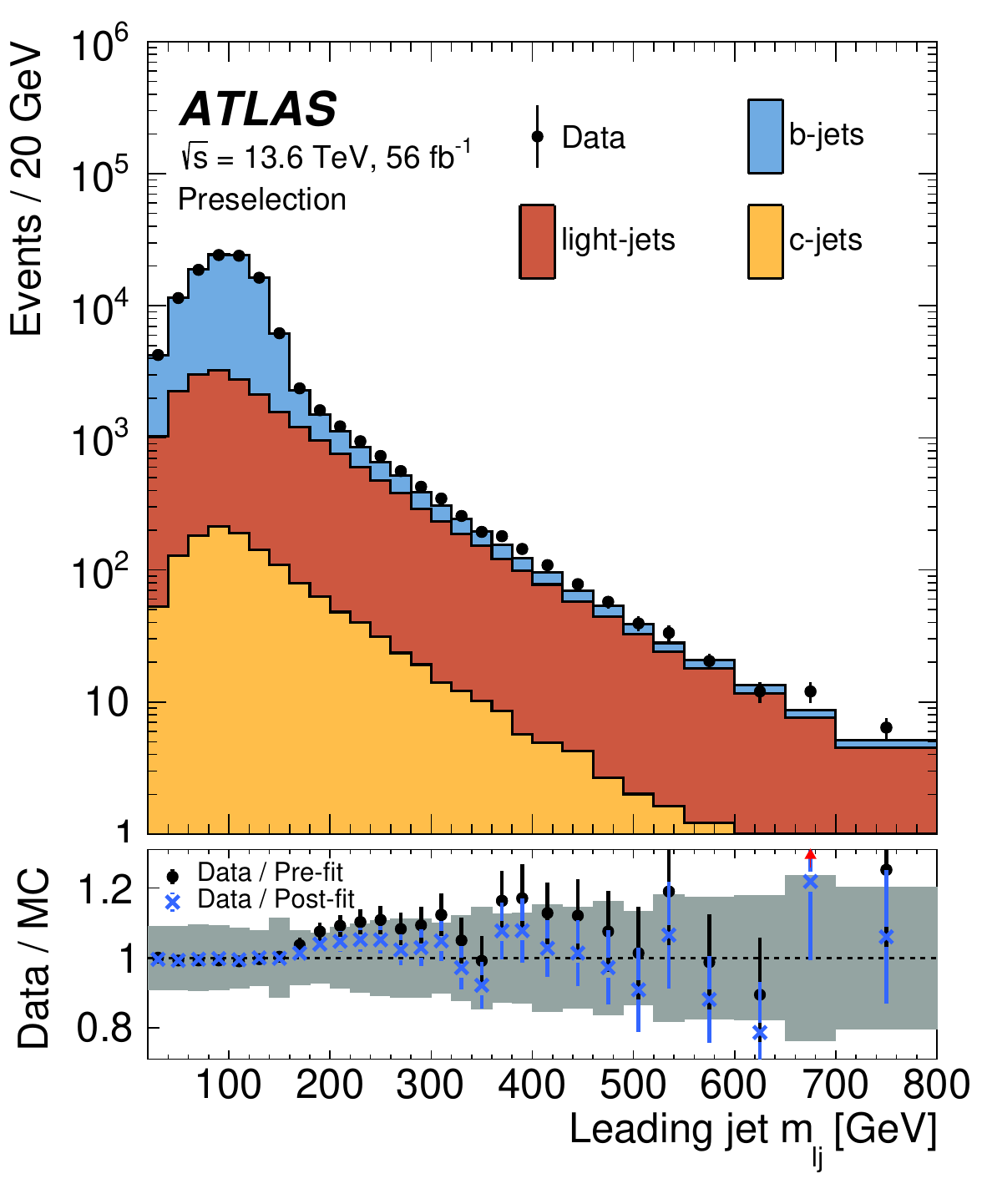}}
\subfloat[]{ \includegraphics[width=0.47\textwidth]{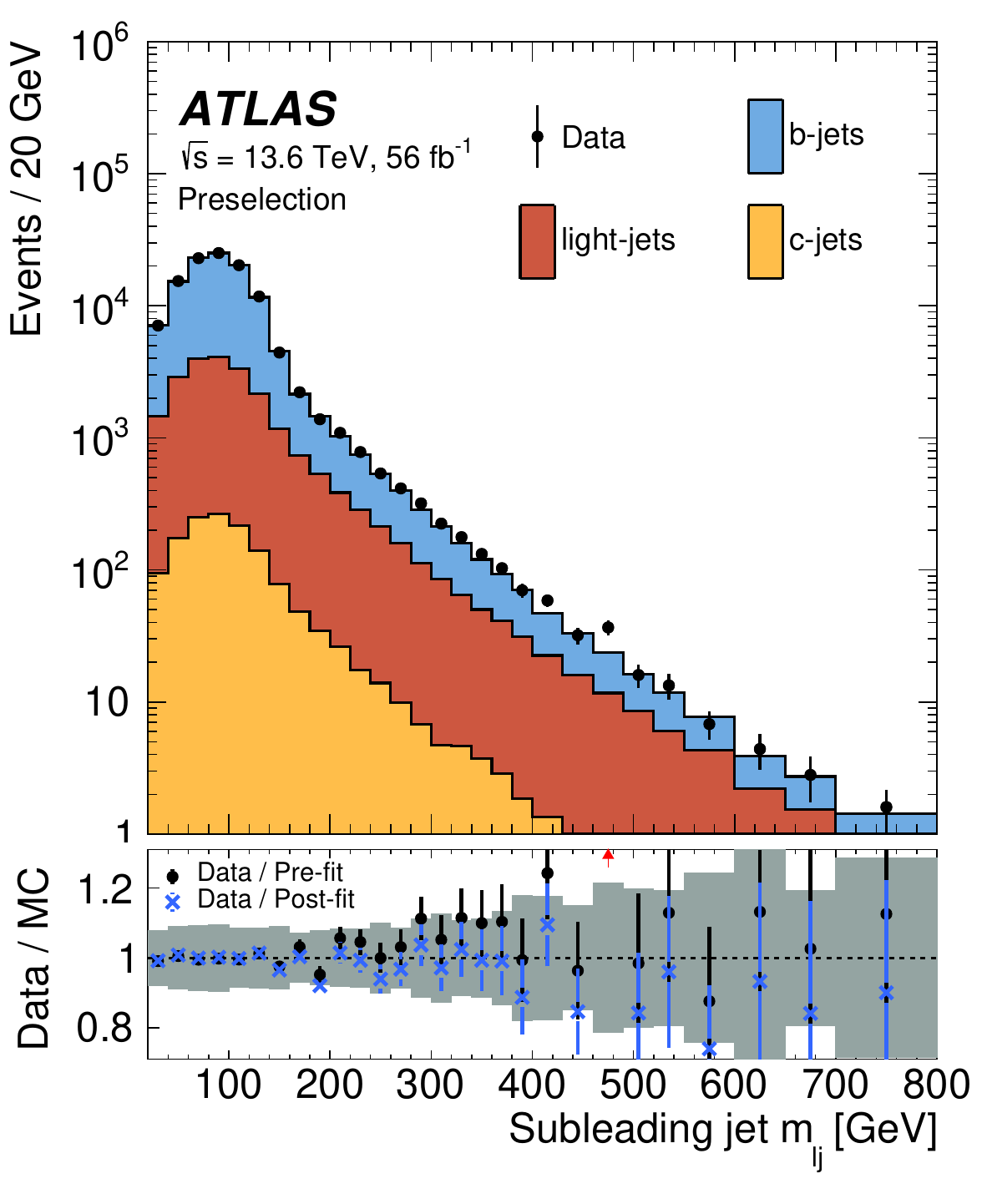}}
\caption{Distribution of the jet--lepton invariant masses for the (a) leading and (b) sub-leading selected jets, shown both in data and simulation after preselection. Simulated events are classified in terms of the flavour composition of the two jets. The non-\bjet contribution in the MC simulation has been corrected by the measured light-jet mis-tagging scale factors, as described in Section~\ref{sec:uncertainties}. The total contribution of simulated MC events is normalised to the total data yield. The bottom panels show the data-to-MC ratio, where the black markers correspond to the pre-fit MC simulation and the blue markers correspond to the post-fit MC simulation. The error bars represent the statistical uncertainty of the data. The shaded band is the sum in quadrature of MC statistical and systematic uncertainties in the pre-fit MC simulation (consisting of detector uncertainties and physics modelling uncertainties described in Section~\ref{sec:uncertainties}).}
\label{fig:SELECTION-mjl-distributions}
\end{figure}

\begin{table}[ht!]
\centering
\caption{Definition of signal and control regions for the simultaneous estimate of the \bjet tagging efficiencies and the jet truth flavour compositions $bb$, $ll$, $bl$, $lb$. The jet--lepton invariant mass is defined as $m_{j,\ell}$.}
\begin{tabular}{ccc}
\toprule
Analysis region & Description & Selection \\
\midrule
SR              & $bb$-enriched & $m_{j1,\ell} < 175$~\GeV and $m_{j2,\ell} < 175$~\GeV \\
CR$_\text{LL}$   & $ll$-enriched & $m_{j1,\ell} \ge 175$~\GeV and $m_{j2,\ell} \ge 175$~\GeV \\
CR$_\text{BL}$   & $bl$-enriched & $m_{j1,\ell} < 175$~\GeV and $m_{j2,\ell} \ge 175$~\GeV \\
CR$_\text{LB}$   & $lb$-enriched & $m_{j1,\ell} \ge 175$~\GeV and $m_{j2,\ell}<175$~\GeV \\
\bottomrule
\end{tabular}
\label{tab:SELECTION-analysis-region-definition}
\end{table}

\begin{figure}
\centering
\includegraphics[width=0.6\textwidth]{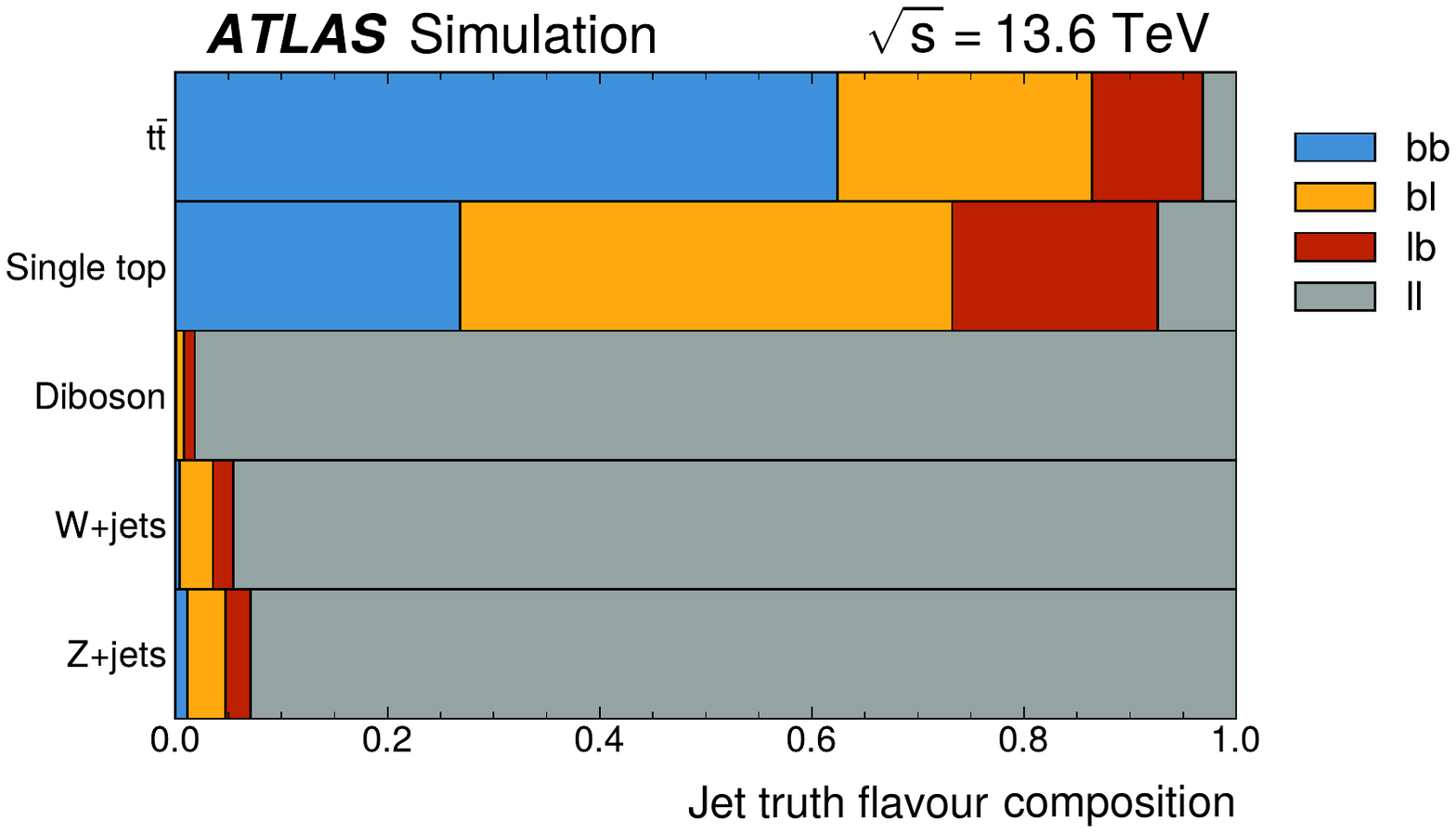}
\caption{Jet truth flavour compositions of \ttbar events and non-\ttbar processes in the SR according to the MC simulation.}
\label{fig:SELECTION-true-flavour-fractions}
\end{figure}

The leading jet \pt distribution in the SR is shown in Figure~\ref{fig:leading_jet_pt}, showing good data--MC agreement within the total uncertainty. The GN2 score distribution of the leading jet is shown in Figure~\ref{fig:leading_jet_gn2} with the six PCBT bins: [100\%, 90\%], [90\%, 85\%], [85\%, 77\%], [77\%, 70\%], [70\%, 65\%], and [65\%, 0\%]. Each PCBT bin represents an exclusive interval of the corresponding GN2 score, or equivalently, an interval of the cumulative \bjet tagging efficiency.

\begin{figure}
\centering
\subfloat[]{
\includegraphics[width=0.45\textwidth]{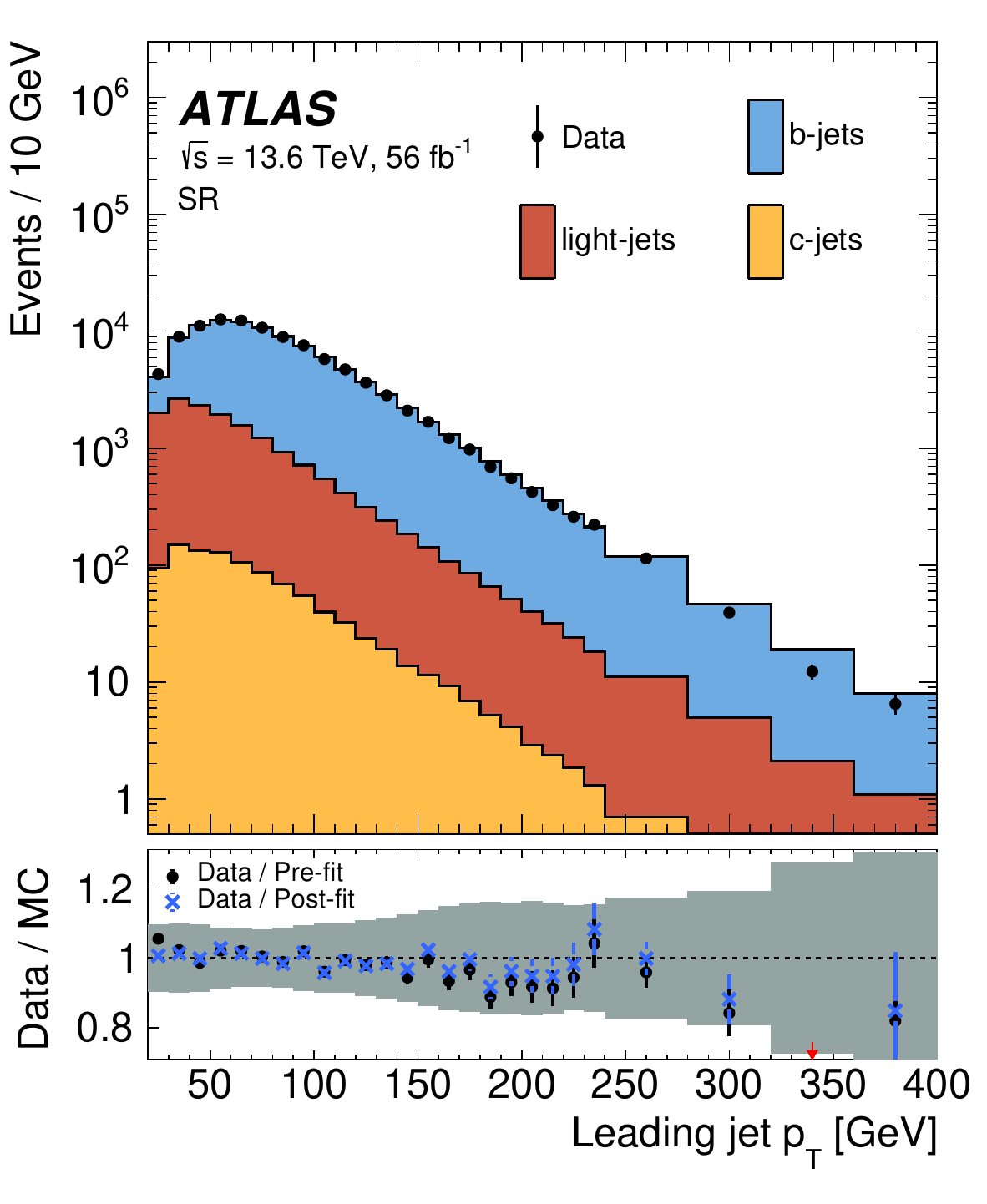}
\label{fig:leading_jet_pt}
}
\subfloat[]{
\includegraphics[width=0.45\textwidth]{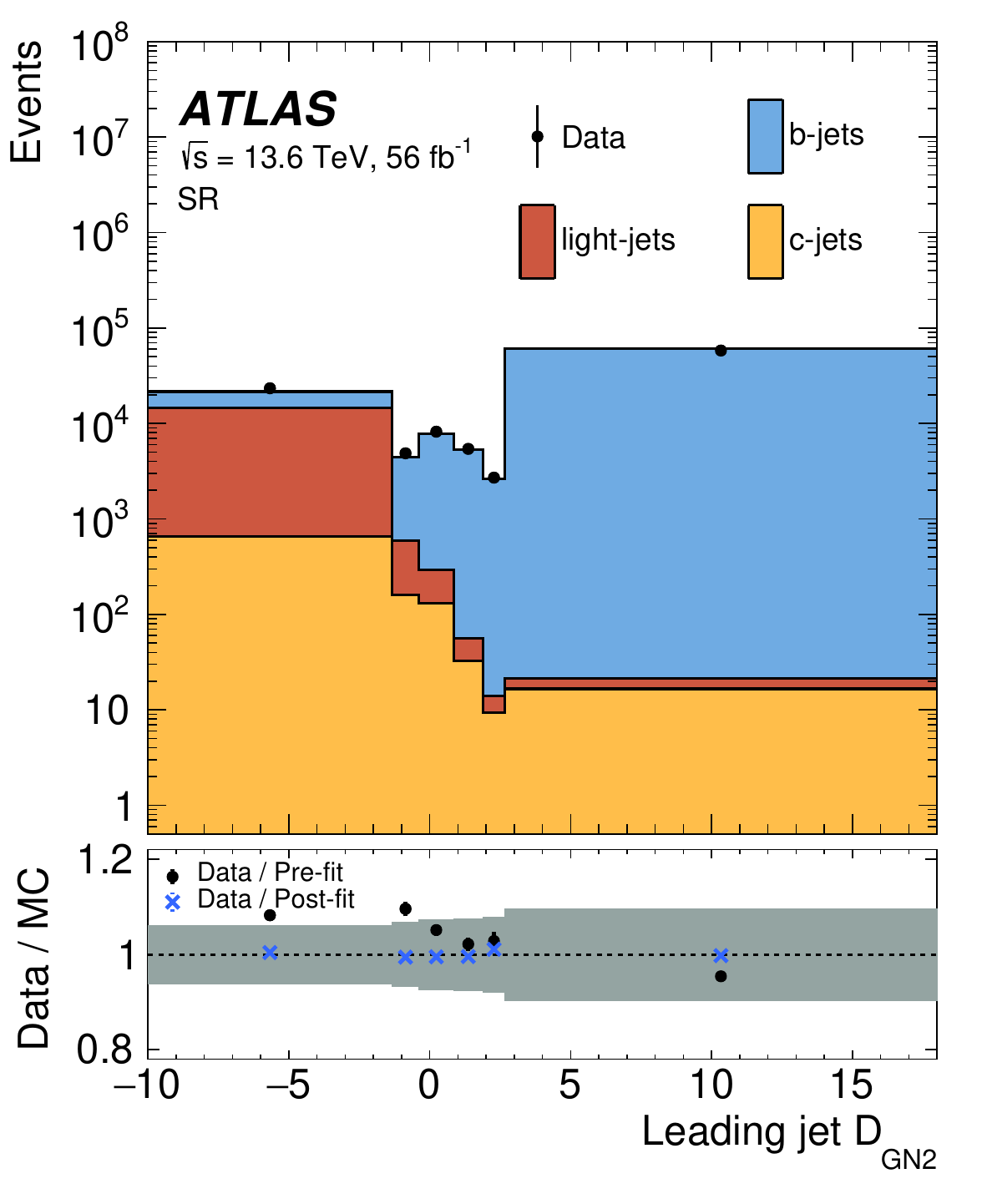}
\label{fig:leading_jet_gn2}
}
\caption{Leading jet (a) \pT distribution and (b) GN2 score distribution within the SR, where the six PCBT bins follow the increase of the GN2 score: [100\%, 90\%], [90\%, 85\%], [85\%, 77\%], [77\%, 70\%], [70\%, 65\%] and [65\%, 0\%]. The non-\bjet contribution in the MC simulation has been corrected by the measured light-jet mis-tagging scale factors, as described in Section~\ref{sec:uncertainties}. The total contribution of simulated MC events is normalised to the total data yield. The bottom panel shows the data-to-MC ratio, where the black markers correspond to the pre-fit MC simulation and the blue markers correspond to the post-fit MC simulation. The error bars represent the data's statistical uncertainty. The shaded band is the sum in quadrature of MC statistical and systematic uncertainties in the pre-fit MC simulation (consisting of detector uncertainties and physics modelling uncertainties described in Section~\ref{sec:uncertainties}).
}
\end{figure}

%

%
%
%
%
%
%
%
%
%
%
%
%
%
%


%
%
%
%

%

%
%
%
%

%

%
\section{Extraction of the $b$-jet tagging efficiency}
\label{sec:fit}
%

The method of \bjet tagging efficiency extraction based on the maximum likelihood approach has been documented in detail in Ref.~\cite{FTAG-2018-01} and is briefly summarised below.

In both the SR and three CRs, the selected events are binned into a two-dimensional grid of the leading jet \pt ($T^m$) and subleading jet \pt ($T^n$). The bin boundaries of [20, 40, 60, 140, 400]~\GeV are selected to have enough number of events to perform the \bjet tagging efficiency measurement in each bin and to ensure the convergence of the likelihood fit. This corresponds to 10 ($T^m$, $T^n$) bins in the SR and in each of the three CRs. The composition of jet flavour is dependent on the jet \pt. For example, the fraction of $bb$ events in the SR as a function of the leading and subleading jet \pT is shown in Figure~\ref{fig:SR-bb-fraction-allMC}. The lower fraction of $bb$ events in the low jet \pt region is due to a lower purity of \ttbar events in this phase space, as compared with the region of $60<\textrm{jet}~\pt<400~\GeV$. In contrast, the high jet \pt area is dominated by the \ttbar process; therefore ensuring fractions of $bb$ events to be higher than 80\%. In the SR, events in each ($T^m$, $T^n$) bin are further divided into $6\times6$ PCBT bins of the leading jet ($O^k$) and subleading jet ($O^p$). This categorisation strategy results in 390 bins in the input to the likelihood fit, including 10 ($T^m$, $T^n$) in each of the CRs and 360 ($T^m, T^n, O^k, O^p$) bins in the SR.

The total prediction of each ($T^m$, $T^n$) bin in one of the CRs (CR$_\text{LL}$, CR$_\text{BL}$, or CR$_\text{LB}$) is expressed as:
\begin{equation*}
\nu_{CR}(T^m, T^n) = c_{bb}^{m,n}\nu_{CR,bb}^{m,n} + c_{bl}^{m,n}\nu_{CR,bl}^{m,n} + c_{lb}^{m,n}\nu_{CR,lb}^{m,n} + c_{ll}^{m,n}\nu_{CR,ll}^{m,n},
\end{equation*}
that is the sum of the predicted numbers of events $\nu$ in all jet flavour compositions, each multiplied by the corresponding jet flavour composition correction factor ($c_{bb}, c_{bl},c_{lb}$ or $c_{ll}$). The jet flavour composition correction factors are shared between the CRs and SR predictions and allow to adjust the jet flavour composition to data. They are determined for each systematic uncertainty variation, which reduces the sensitivity to the mis-modelling of jet flavour composition in the simulation and improve the precision of the measurement. The total prediction of each ($T^m, T^n, O^k, O^p$) bin in the SR is written as:
\begin{align*}
\nu_{SR}(T^m, T^n, O^k, O^p)  &= c_{bb}^{m,n}\nu_{SR,bb}^{m,n}\cdot\mathcal{P}_b(O^k|T^m)\cdot\mathcal{P}_b(O^p|T^n) \nonumber\\
&+ c_{bl}^{m,n}\nu_{SR,bl}^{m,n}\cdot\mathcal{P}_b(O^k|T^m)\cdot\mathcal{P}_l(O^p|T^n) \nonumber\\
&+ c_{lb}^{m,n}\nu_{SR,lb}^{m,n}\cdot\mathcal{P}_l(O^k|T^m)\cdot\mathcal{P}_b(O^p|T^n) \nonumber\\
&+ c_{ll}^{m,n}\nu_{SR,ll}^{m,n}\cdot\mathcal{P}_l(O^k|T^m)\cdot\mathcal{P}_l(O^p|T^n),
\end{align*}
where the $\mathcal{P}_b$ are the \bjet tagging efficiencies to be measured in PCBT bins for each interval of jet \pt and $\mathcal{P}_l$ are the light-jet mis-tagging rates computed from the MC, where the correction on the light-jet mis-tagging rate derived from a dedicated calibration analysis has been accounted for and applied to the total non-\bjet contribution.

The likelihood is defined as:
\begin{equation}
\ln \mathcal{L} (\hat \Theta) = \sum_{m,n} \left[ \sum_{\text{LL, BL, LB}} \left( - \nu_{CR} (\hat \Theta) + n_{CR} \ln \nu_{CR} (\hat \Theta) \right) + \sum_{k,p} \left( -\nu_{SR} (\hat \Theta) + n_{SR} \ln \nu_{SR} (\hat \Theta) \right) \right],
\end{equation}
where $n_{CR}$ is the observed number of events in each $(T^m,\ T^n)$ bin for the three CRs, $n_{SR}$ is that in each $(O^k, O^p, T^m, T^n)$ bin for the SR, and
\begin{equation*}
\hat \Theta = \left( \mathcal{P}_b(O^k|T^m), c_{bb}^{m,n}, c_{bl}^{m,n}, c_{lb}^{m,n}, c_{ll}^{m,n} \right)
\end{equation*}
is the list of free parameters to be determined. The unitarity constraint is applied to the \bjet tagging efficiency from all PCBT bins, resulting in five free parameters of $\mathcal{P}_b$ in each of the four jet \pt bins. In total, there are fourty jet flavour composition correction factors and twenty $\mathcal{P}_b$'s to be determined by the fit. The minimisation of the likelihood with respect to the data, which is evaluated numerically by RooFit~\cite{RooFit}, returns the most probable value of $\hat \Theta$ and their statistical uncertainty due to the finite number of events in the data.

\begin{figure}
\centering
\includegraphics[width=0.60\textwidth]{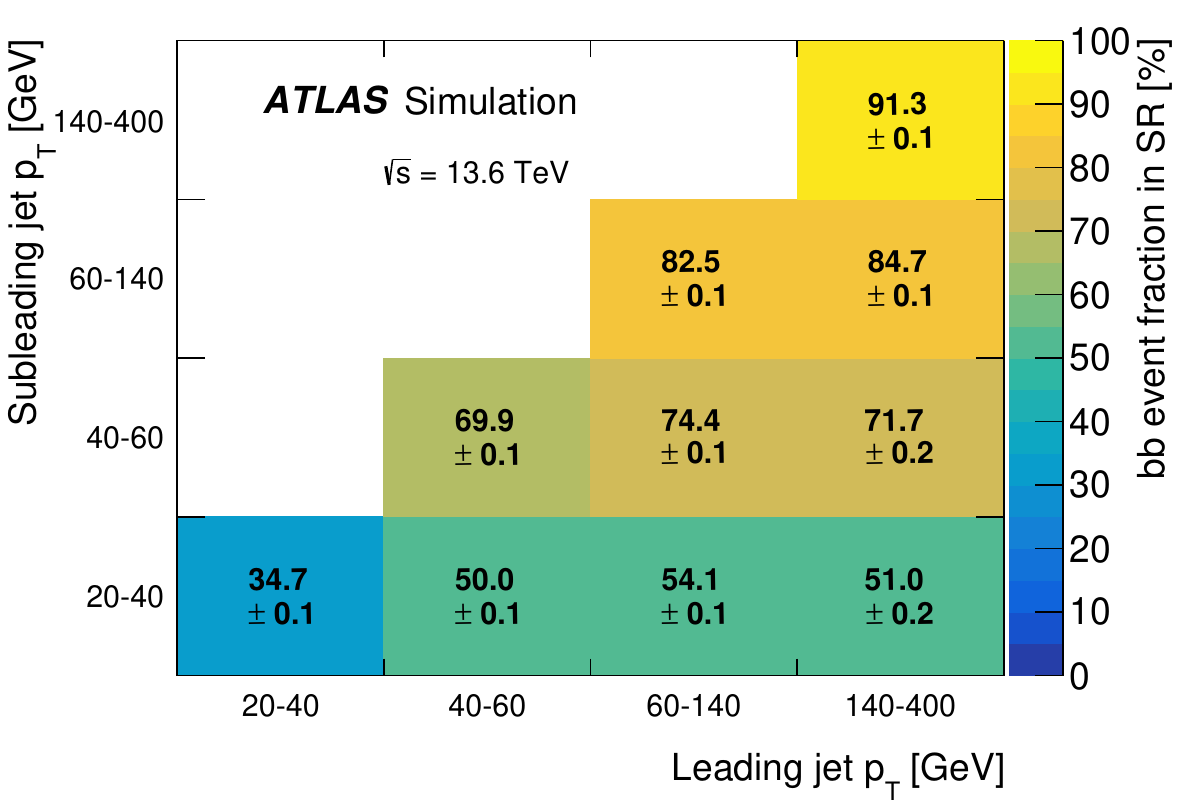}
\caption{Fraction of $bb$ flavour events in the signal region from all simulated MC contributions as a function of the leading and subleading jet \pT. Uncertainties are statistical only.}
\label{fig:SR-bb-fraction-allMC}
\end{figure}

The likelihood fit has been tested for bias using pseudo-data from the nominal MC simulation and from MC samples with varied jet flavour compositions. These tests showed the fit introduces negligible bias, so no additional systematic uncertainty related to the fitting procedure is required.


%
\section{Uncertainties}
\label{sec:uncertainties}
%

Uncertainties related to the measurement of the \bjet tagging efficiency arise from both statistical and systematic origin due to detector effects and physics modelling.

The data statistical uncertainty in the \bjet tagging efficiency is directly taken from the error matrix returned by the minimisation of the likelihood. The size of the data statistical uncertainty varies from 0.5\% in the [65\%, 0\%] PCBT bin and medium jet \pt range of $60<\pt<140~\GeV$ to around 10\% in the [100\%, 90\%] PCBT bin and low jet~\pt range of $20<\pt<40~\GeV$. The uncertainty in the \bjet tagging efficiency due to the finite number of simulated MC events of all processes is estimated by the bootstrap method~\cite{ATL-PHYS-PUB-2021-011}. The standard deviation of \bjet tagging efficiency obtained from repeated fits using resampled MC events is taken to be the MC statistical uncertainty. This uncertainty source is the largest in the [100\%, 90\%] PCBT bin and low jet~\pt range of $20<\pt<40~\GeV$, where it reaches 3\%. In the other PCBT bins, the MC statistical uncertainty amounts to no more than 0.5\%.

Systematic uncertainties in the \bjet tagging efficiency are estimated by replacing the nominal MC prediction with a systematic variation and repeating the likelihood fit. The offset between the efficiency from the nominal fit to the data and the efficiency obtained with the systematic variation is taken to be the systematic uncertainty of the source under study. The uncertainties of up and down variations of the same systematic source on the \bjet tagging efficiency are symmetrised with respect to the result of the nominal fit. The first group of systematic uncertainties from the detector response is related to the uncertainty in the lepton selection efficiency, the lepton energy correction, energy scale and resolution of jets, the uncertainty in pile-up modelling and the uncertainty related to the performance of the JVT algorithm~\cite{ATLAS-CONF-2014-018}. Jet energy scale and resolution uncertainties are the leading uncertainty sources related to detector response. The detector uncertainties are below 4\% in all bins except the [100\%, 90\%] PCBT bin in the low jet \pt range of $20<\pt<40~\GeV$, where it can go up to 18\% due to large jet energy scale and resolution uncertainties. The second group of systematic uncertainties concerns the physics modelling of \ttbar and single top-quark production. This group of systematic uncertainties consists of the following:
\begin{itemize}
\item Uncertainties originating from PDFs are accounted for by following the PDF4LHC recommendations~\cite{Butterworth:2015oua,Ball:2022oua}. The five-flavour scheme with variation on \alphas is used for the \ttbar production process and the single top production process in the $s$- and $Wt$ channels, while the four-flavour scheme is used for the single top production in the $t$-channel. The PDF uncertainty is considered uncorrelated between different processes.
\item Uncertainties in the QCD radiation are evaluated by increasing or decreasing the amount of initial-state radiation (ISR) and final-state radiation (FSR) within \PYTHIA 8~\cite{ATL-PHYS-PUB-2016-020}. These uncertainties are considered uncorrelated between different processes.
\item Uncertainties in the matching between the matrix element calculation and the parton shower are taken into account by changing the choice of the \POWHEG-hardness definition and using a different damping factor. The MC sample of this variation is described earlier in Section~\ref{sec:DataMC}.
\item Uncertainties in the modelling of the parton shower and hadronisation processes are assessed by comparing the difference between \PYTHIA 8 and \HERWIG 7.
\item The impact of the interference between the single top $tW$ and \ttbar production is investigated by comparing the difference between the nominal ``diagram removal'' scheme and the alternative ``diagram subtraction'' scheme, as discussed in Section~\ref{sec:DataMC}.
\end{itemize}
The size of the \ttbar modelling uncertainties ranges from 0.5\% in the [70\%, 65\%] PCBT bin and medium \pt region of $60<\pt<140~\GeV$ to 14\% in the [100\%, 90\%] PCBT bin and low jet \pt region of $20<\pt<40~\GeV$, where the uncertainties in the modelling of parton shower and hadronisation processes and in the modelling of QCD FSR dominate. The single top modelling uncertainties, dominated by the difference between the 'diagram removal' scheme and the 'diagram subtraction' scheme, are below 1\% except the [100\%, 90\%] PCBT bin. In the [100\%, 90\%] PCBT bin, the single top modelling uncertainties are the largest, where they amount to the maximum of 6\% in the region of $140<\textrm{jet}~\pt<400~\GeV$. The other sources of physics modelling uncertainties not listed above, including the uncertainties in the production cross-sections of $V+$jets and diboson processes, were found to be negligible in previous measurement of the \bjet tagging efficiencies~\cite{FTAG-2018-01} and are therefore not included in this work.

The last group of systematic uncertainties consists of two sources estimated from dedicated procedures. The first source is the uncertainty in the rate of electrons not originating from the \Zboson- or \Wboson-boson decay (non-prompt electrons) in the MC simulation. The other uncertainty originates from the correction on the probability for non-\bjets, mainly light-flavour jets, to mistakenly be identified as \bjets.
Non-prompt leptons are included in the simulation of the nominal MC. Even though non-prompt electrons predicted by the simulation only reach per-mille level fraction in the total event yield, the mis-modelling of the non-prompt electron rate is expected to introduce a non-negligible source of uncertainty to the measurement of the \bjet tagging efficiency. The uncertainty variation on the non-prompt electron rate is derived from a data-driven approach. First, event selections are carried out as described in Section~\ref{sec:object}, but two same-charged leptons are required instead of the nominal opposite-charge selection. Then, the ratio of data subtracted by the MC contribution with a prompt electron over the MC events with a non-prompt electron is computed in three coarse bins of electron \pt: [0, 150, 300, 400]~\GeV. The ratio is the highest for $0<\textrm{electron}~\pt<150~\GeV$, reaching 5.5 (3.4) in the data taken during 2022 (2023). For $300<\textrm{electron}~\pt<400~\GeV$, the ratio drops to 1.3 (1.6) for the data taken during 2022 (2023). Finally, this ratio is applied as correction factors to reweight the events with a non-prompt electron in the nominal MC as a function of electron \pt, thereby producing the uncertainty variation of the non-prompt electron rate.
Typically, the non-prompt electron rate uncertainty is important only in the lowest jet \pt bin of $20<\pt<40~\GeV$, spanning the range from 1\% to 7\%.

The central value of the total non-\bjet mis-tagging rate is corrected using scale factors derived from a dedicated calibration analysis for light-jet mis-tagging rate. The method of the calibration is similar to the one described in~\cite{FTAG-2019-02}. The light-jet mis-tagging scale factors are compatible with unity for the [100\%, 90\%] PCBT bin, and vary between 1.3 and 1.5 most of the other PCBT bins. The impact of the uncertainties associated with the light-jet mis-tagging rate calibration is evaluated in this measurement. In all PCBT bins except the [65\%, 0\%] PCBT bin, the mis-tagging systematic uncertainties in the \bjet tagging efficiency is the largest in the first jet \pt bin between 20 and 40~\GeV. The relative uncertainty amounts to 2\% on the \bjet tagging efficiency in the [100\%, 90\%] and [90\%, 85\%] PCBT bins for $20<\textrm{jet}~\pt<40~\GeV$, where the fraction of non-truth \bjet is the largest. The mis-tagging systematic uncertainties decreases with a higher fraction of \bjet contribution in the PCBT bins.

The total uncertainty is the sum in quadrature of the data statistical uncertainty, the MC statistical uncertainty and all the systematic uncertainties, computed for each PCBT and jet \pt bin and ranges from 1\% in the [65\%, 0\%] PCBT bin for $60<\textrm{jet}~\pt<140~\GeV$ to 25\% in the [100\%, 90\%] PCBT bin for $20<\textrm{jet}~\pt<40~\GeV$.


%
\section{Results}
\label{sec:result}
%

With the likelihood fit described in Section~\ref{sec:fit}, measurements of the GN2 \bjet tagging efficiency are carried out in each PCBT bin as a function of jet \pt, as shown in Figure~\ref{fig:result_Pb_all}. For illustrative purposes, the measurements are also compared with the efficiency evaluated from simulated \ttbar samples after the SR selection. For each PCBT bin $O^k$ and jet \pt interval $T^m$, the simulation-to-data scale factor of the \bjet tagging efficiency is the ratio of the measured \bjet tagging efficiency ($\mathcal{P}^\text{Data}_b(O^k|T^m)$) over the \bjet tagging efficiency in the sum of all nominal MC samples ($\mathcal{P}^\text{MC}_b(O^k|T^m)$)
\begin{equation*}
SF(O^k,T^m) = \frac{\mathcal{P}^\text{Data}_b(O^k|T^m)}{\mathcal{P}^\text{MC}_b(O^k|T^m)}.
\end{equation*}
Scale factors are derived for all PCBT bins as a function of jet \pt, as shown in Figure~\ref{fig:result_SF_all}. The scale factors, ranging from 0.9 to 1.3, are applied as weights in ATLAS physics analyses. The post-fit values of the flavour composition correction factors $c_{bb}$, $c_{bl}$, $c_{lb}$ and $c_{ll}$ range from 0.7 to 1.3, depending on the flavour composition and jet \pt intervals, and are listed in Table~\ref{tab:flavour_component_correction}.
\begin{figure}[tbph]
\centering
\subfloat[]{\includegraphics[width=.42\textwidth]{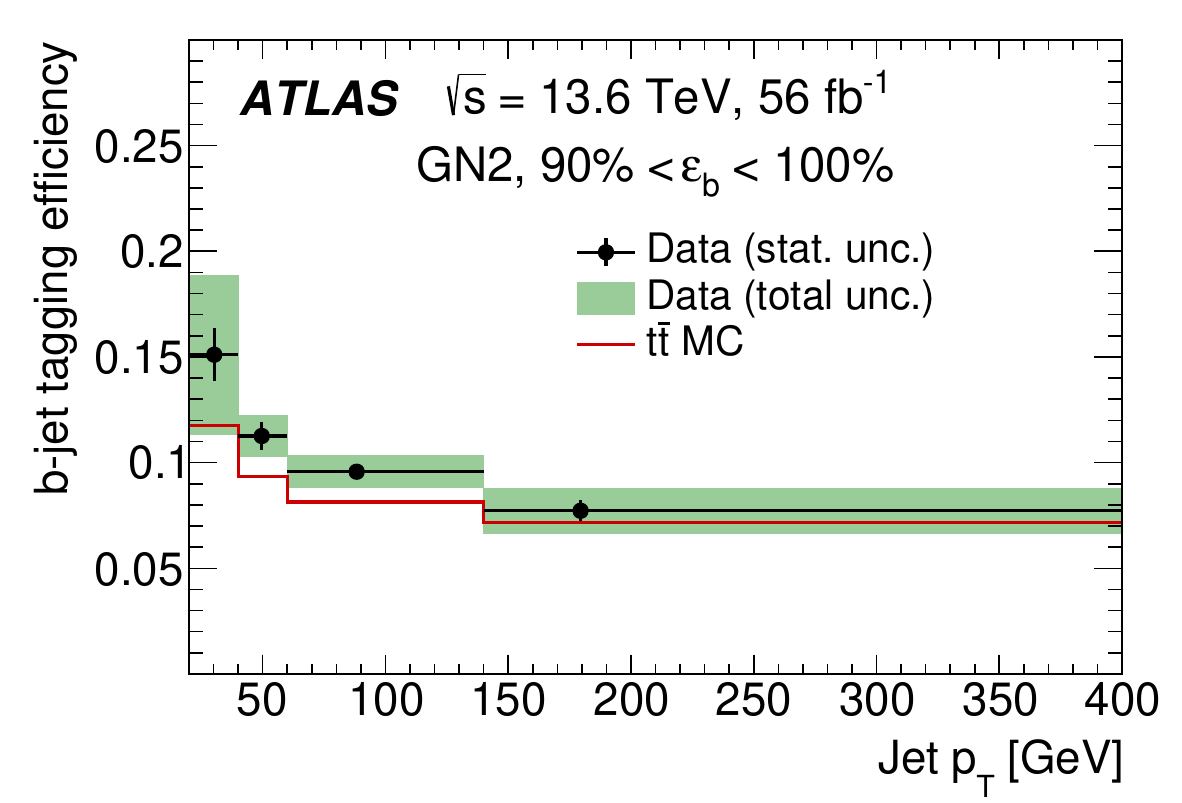}}%
\subfloat[]{\includegraphics[width=.42\textwidth]{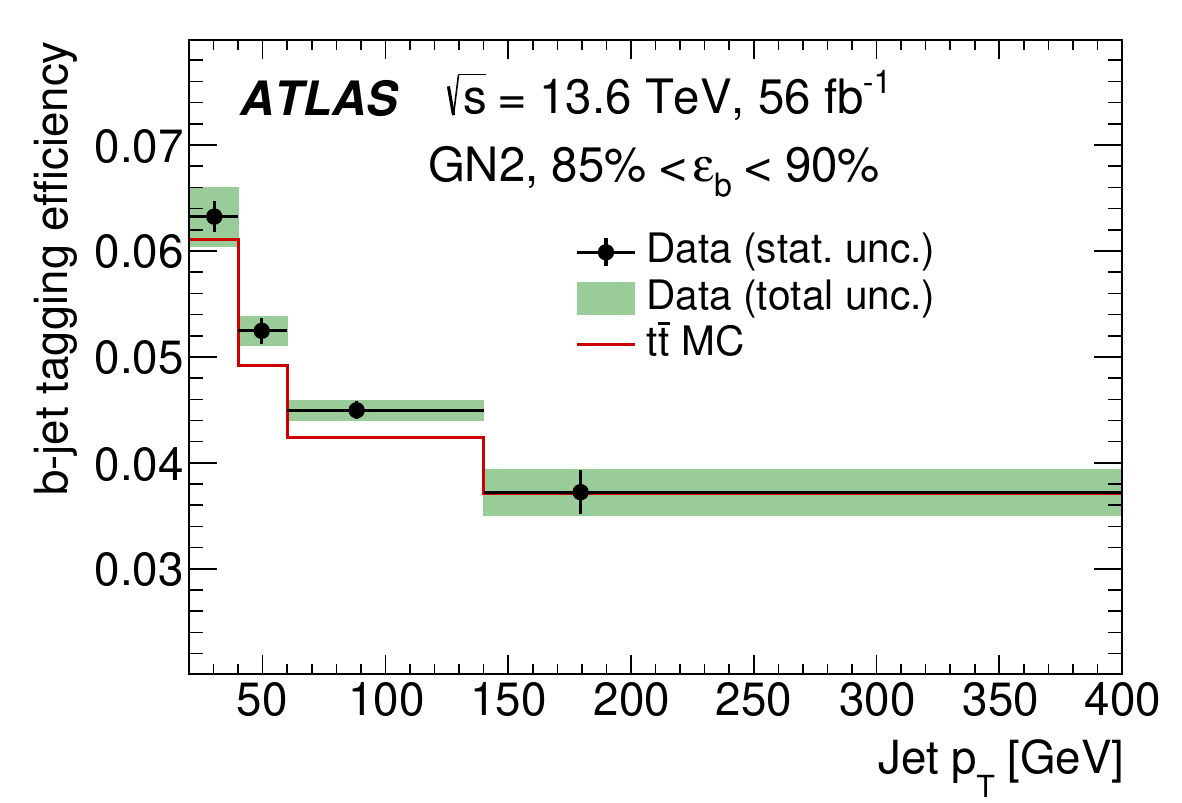}}\\
\subfloat[]{\includegraphics[width=.42\textwidth]{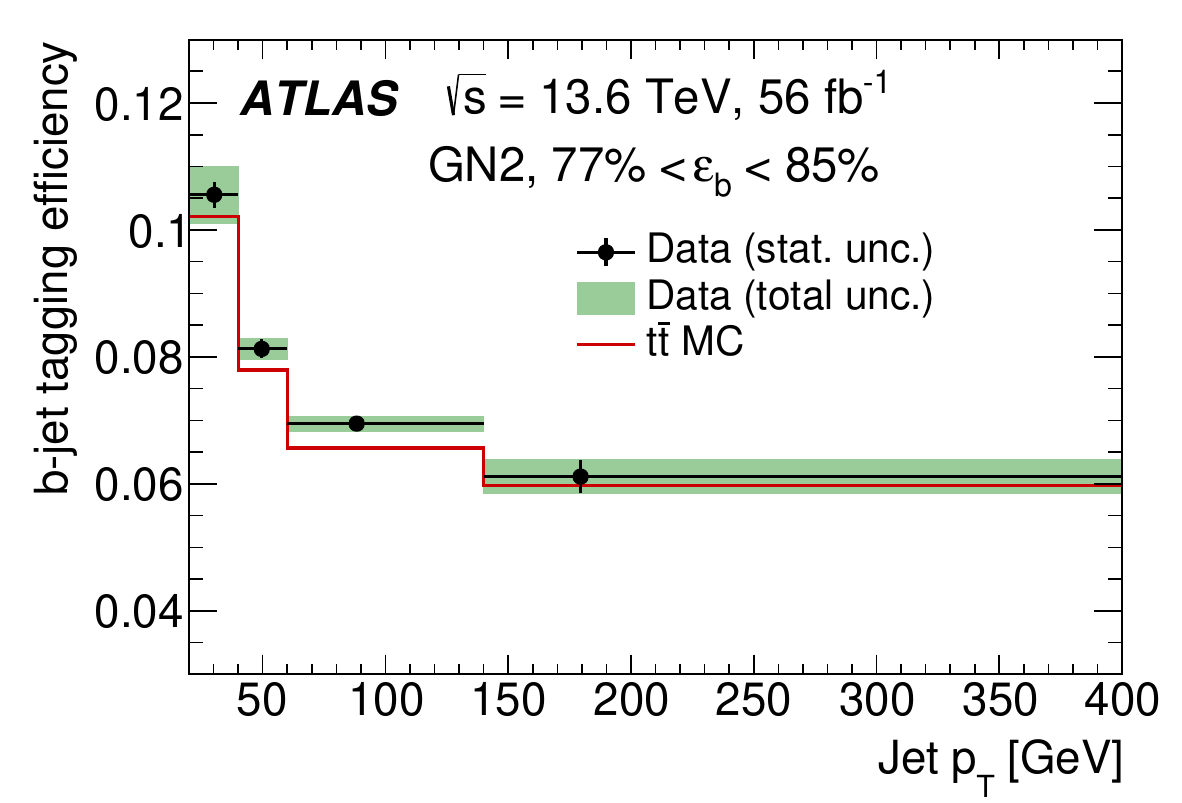}}%
\subfloat[]{\includegraphics[width=.42\textwidth]{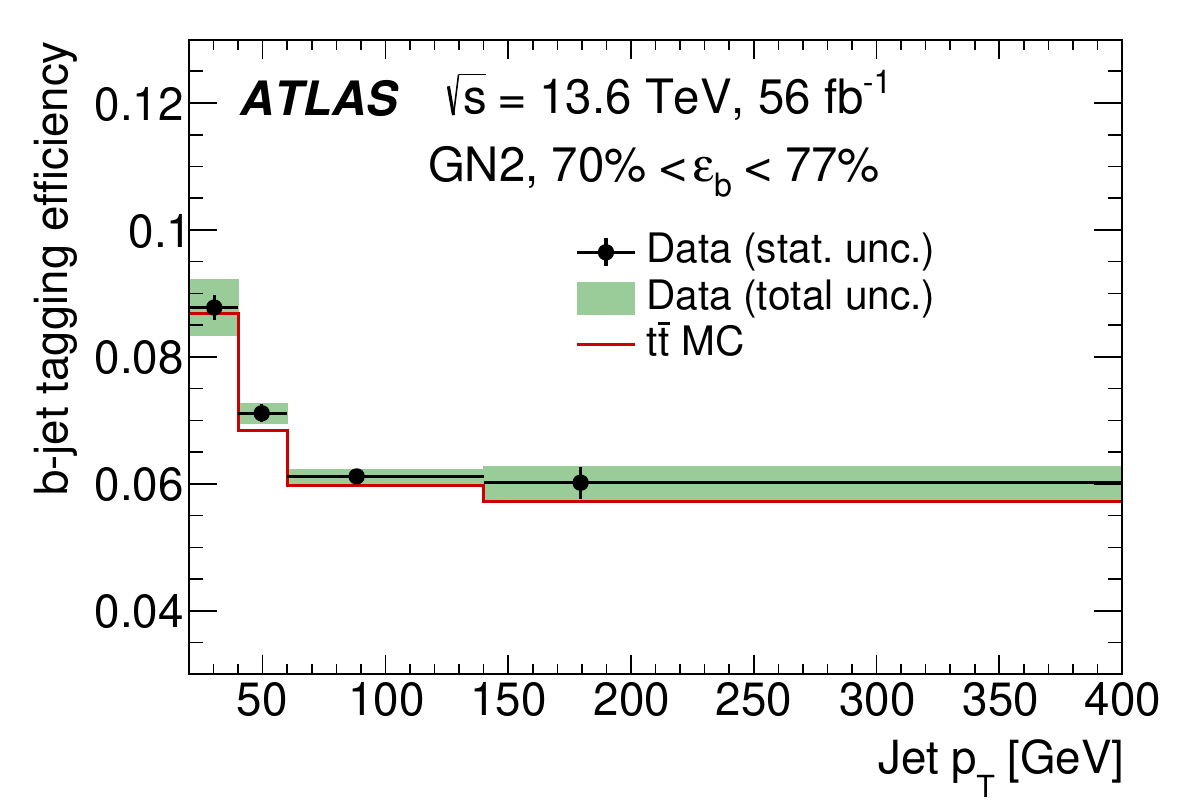}}\\
\subfloat[]{\includegraphics[width=.42\textwidth]{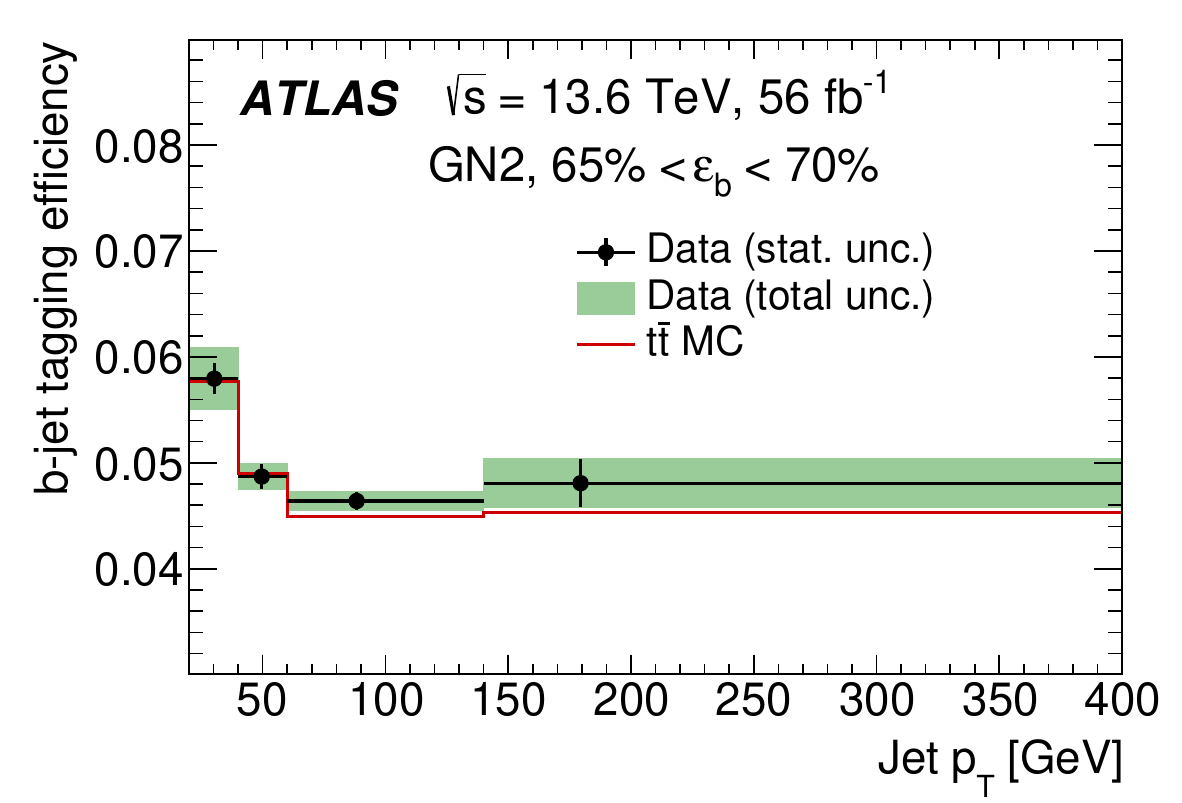}}%
\subfloat[]{\includegraphics[width=.42\textwidth]{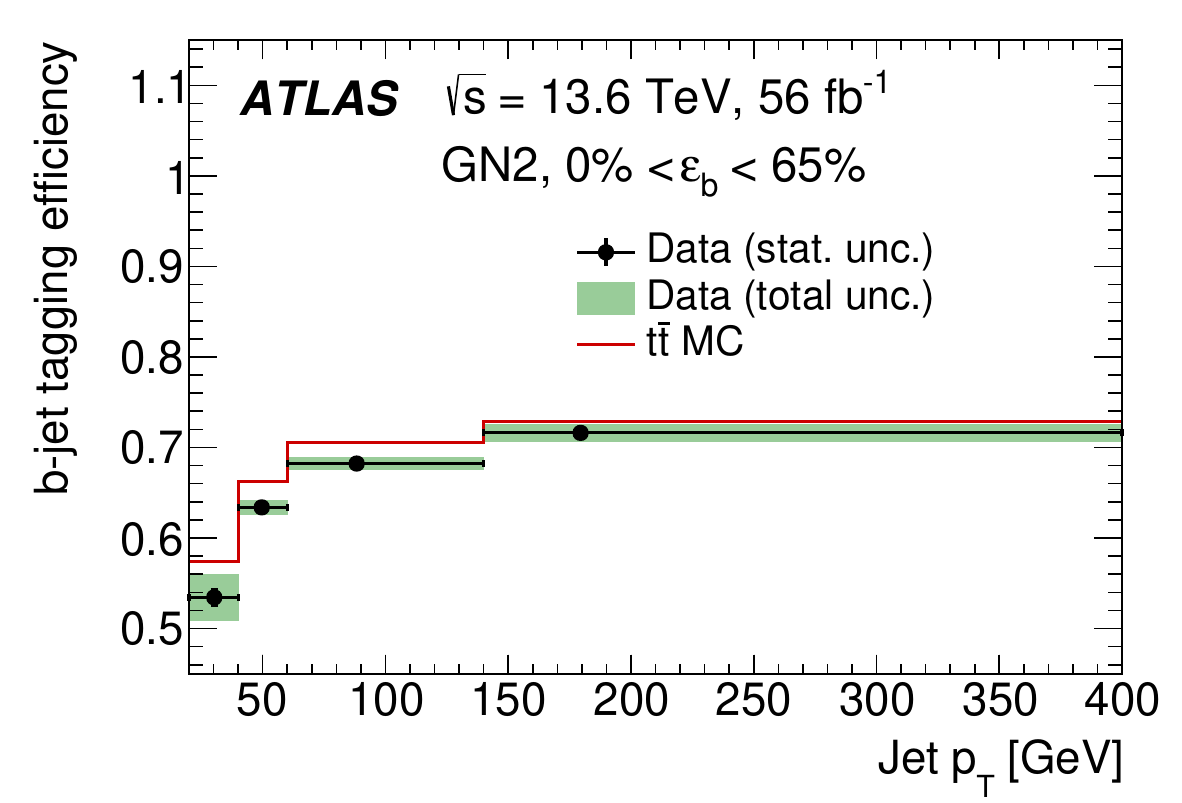}}%
\caption{GN2 \bjet tagging efficiency as a function of jet \pt for all PCBT bins: (a) [100\%, 90\%], (b) [90\%, 85\%], (c) [85\%, 77\%], (d) [77\%, 70\%], (e) [70\%, 65\%] and (f) [65\%, 0\%]. The measurement is shown together with the efficiency derived from simulated \ttbar events satisfying the SR selection. Vertical error bars include data statistical uncertainties only while the green bands correspond to the sum in quadrature of data statistical, MC statistical and systematic uncertainties. The markers are located at the mean of the \bjet \pt distribution in each \pt bin.}
\label{fig:result_Pb_all}
\end{figure}

\begin{figure}[htbp]
\centering
\subfloat[]{
\includegraphics[width=0.42\textwidth]{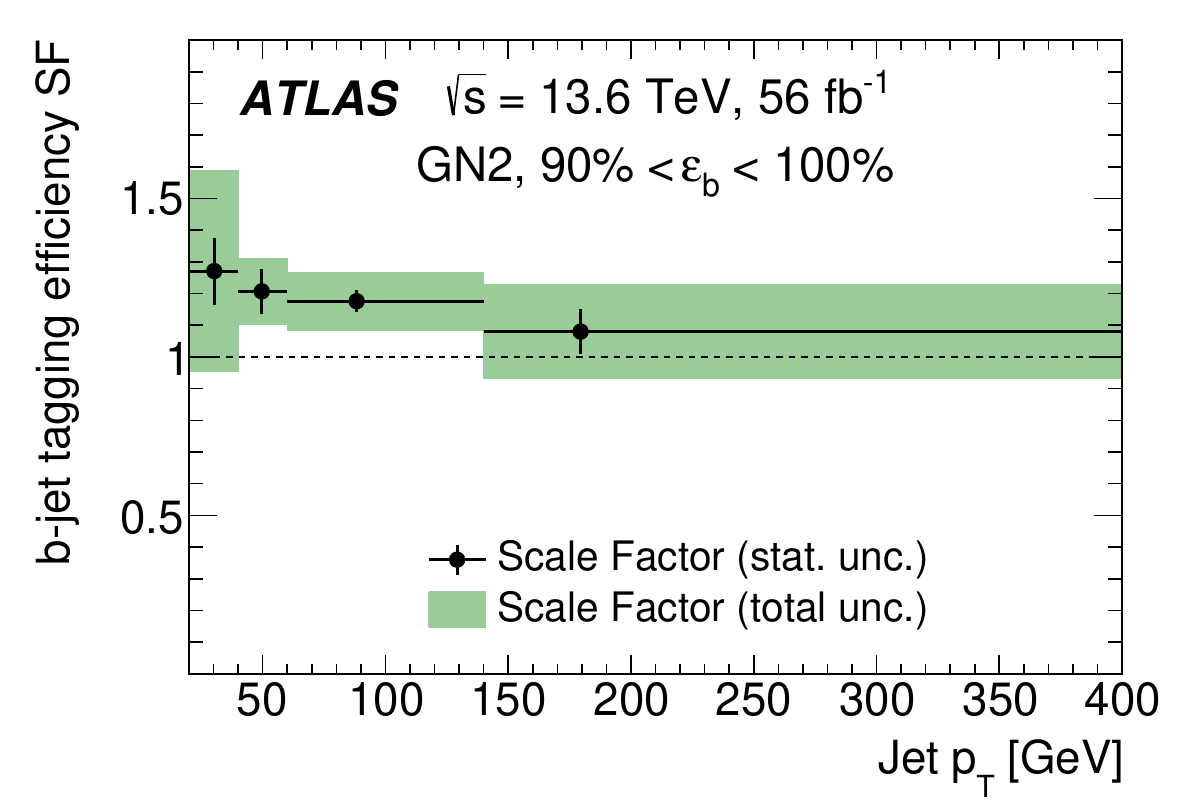}
}
\subfloat[]{
\includegraphics[width=0.42\textwidth]{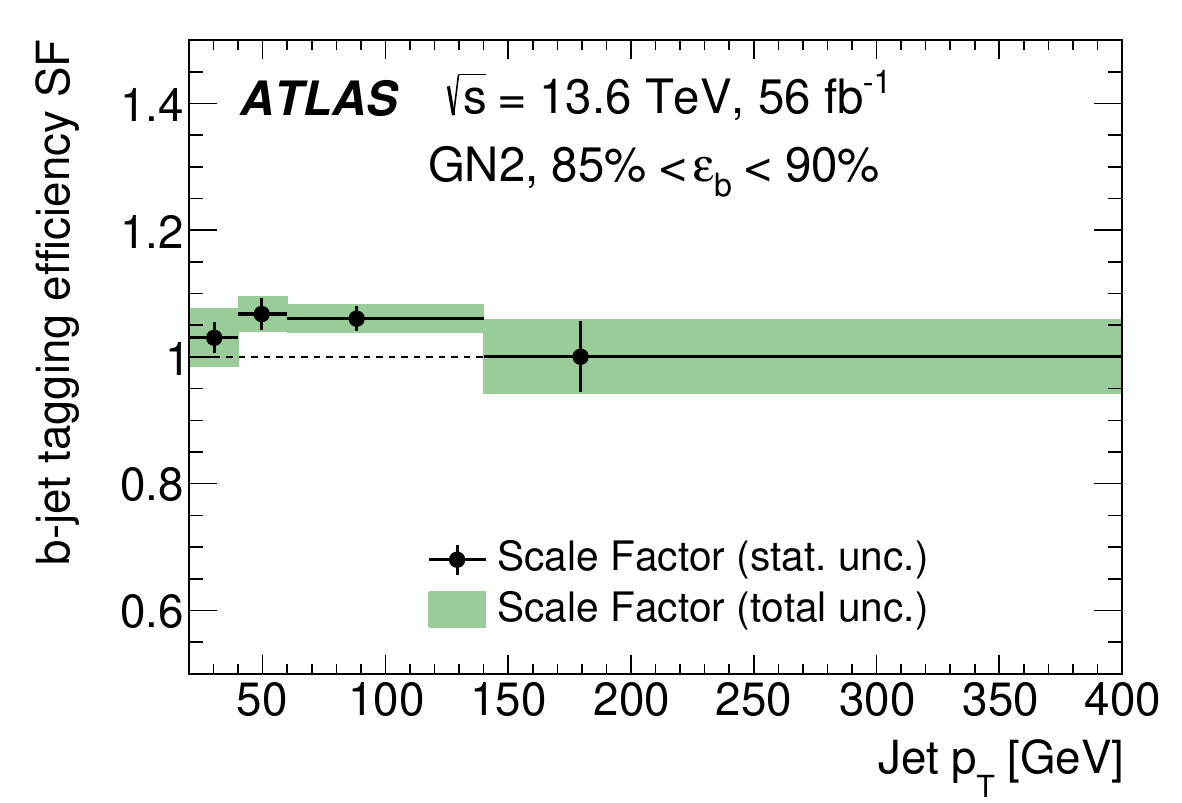}
}\\
\subfloat[]{
\includegraphics[width=0.42\textwidth]{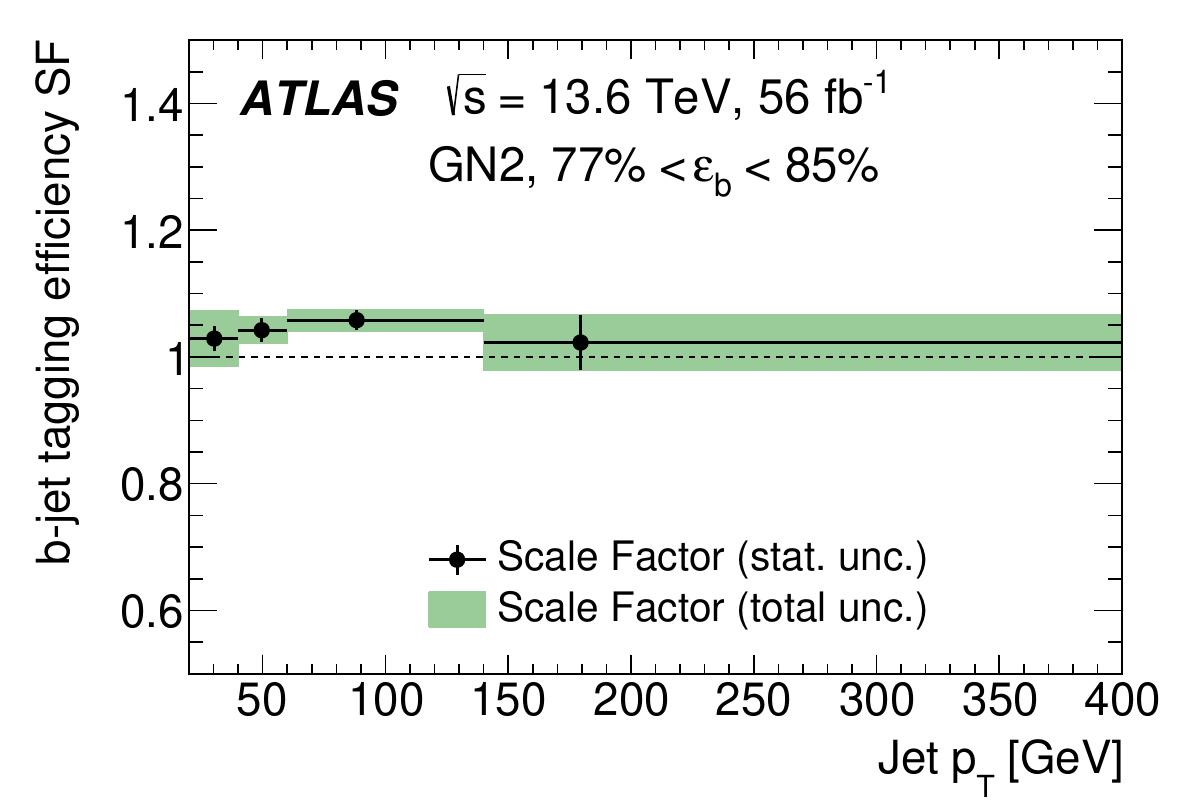}
}
\subfloat[]{
\includegraphics[width=0.42\textwidth]{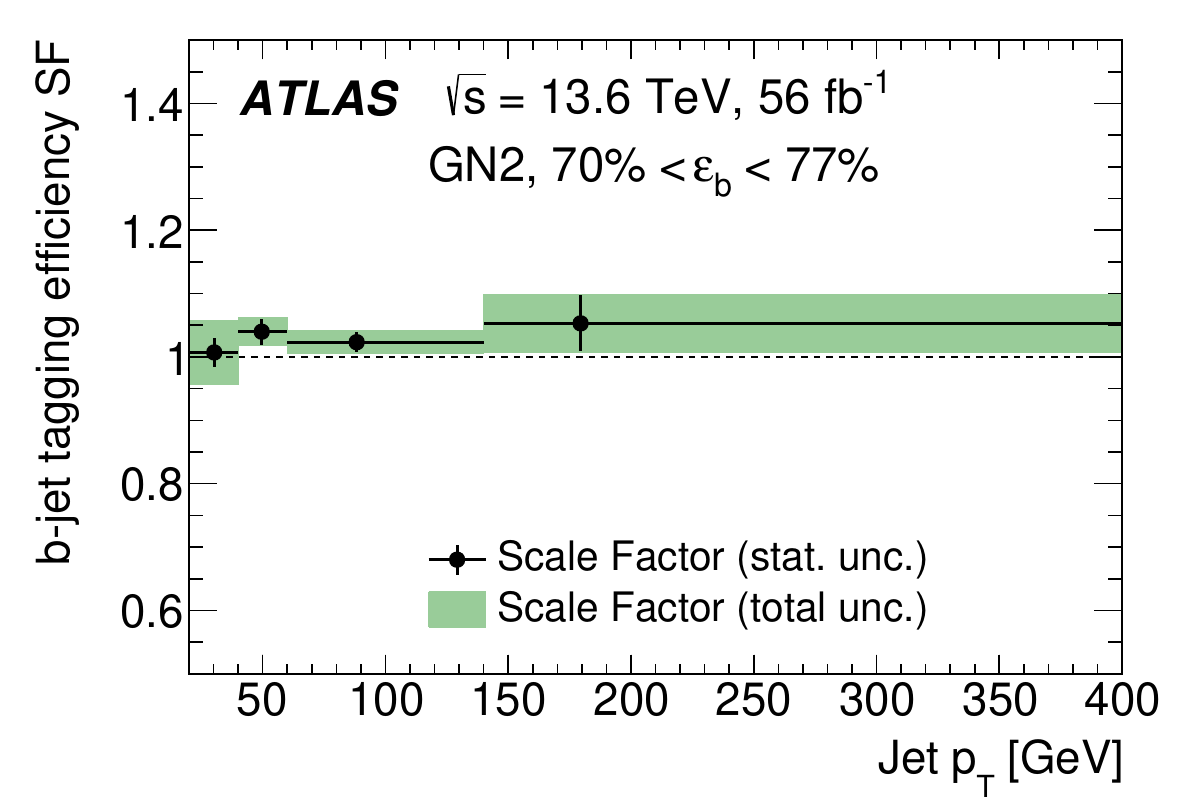}
}\\
\subfloat[]{
\includegraphics[width=0.42\textwidth]{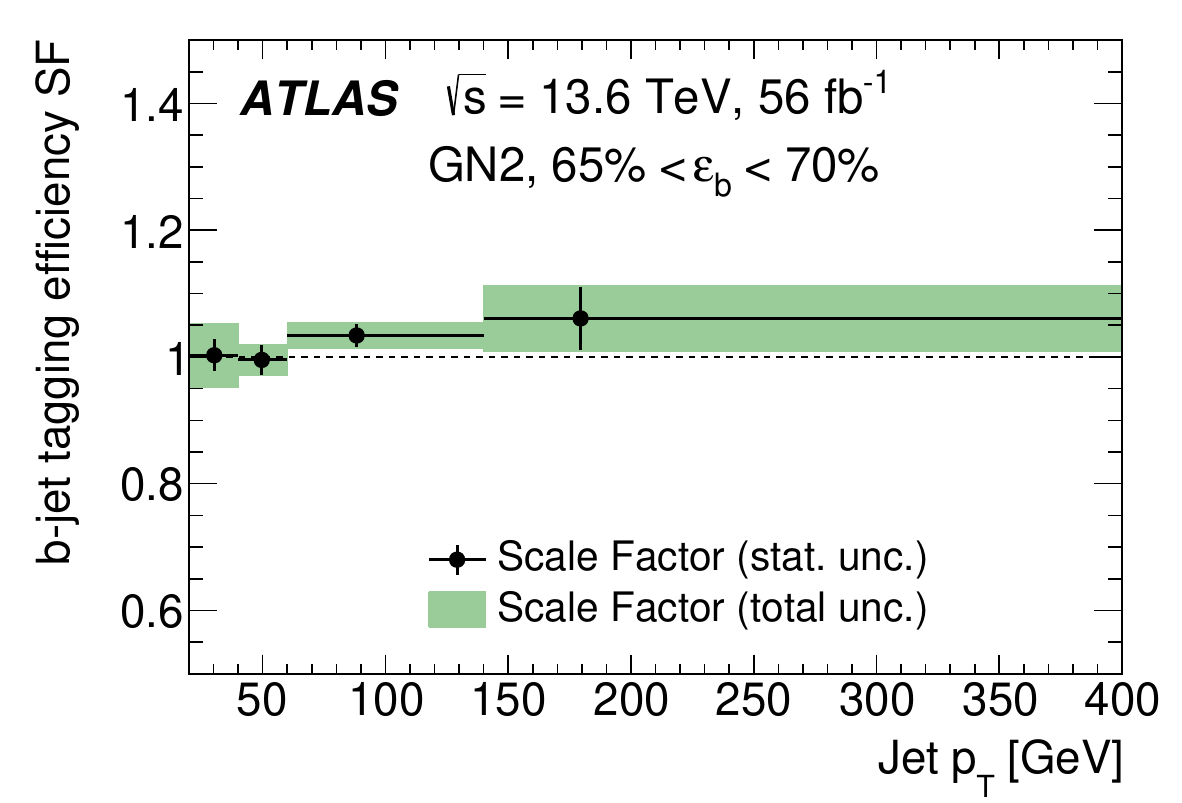}
}
\subfloat[]{
\includegraphics[width=0.42\textwidth]{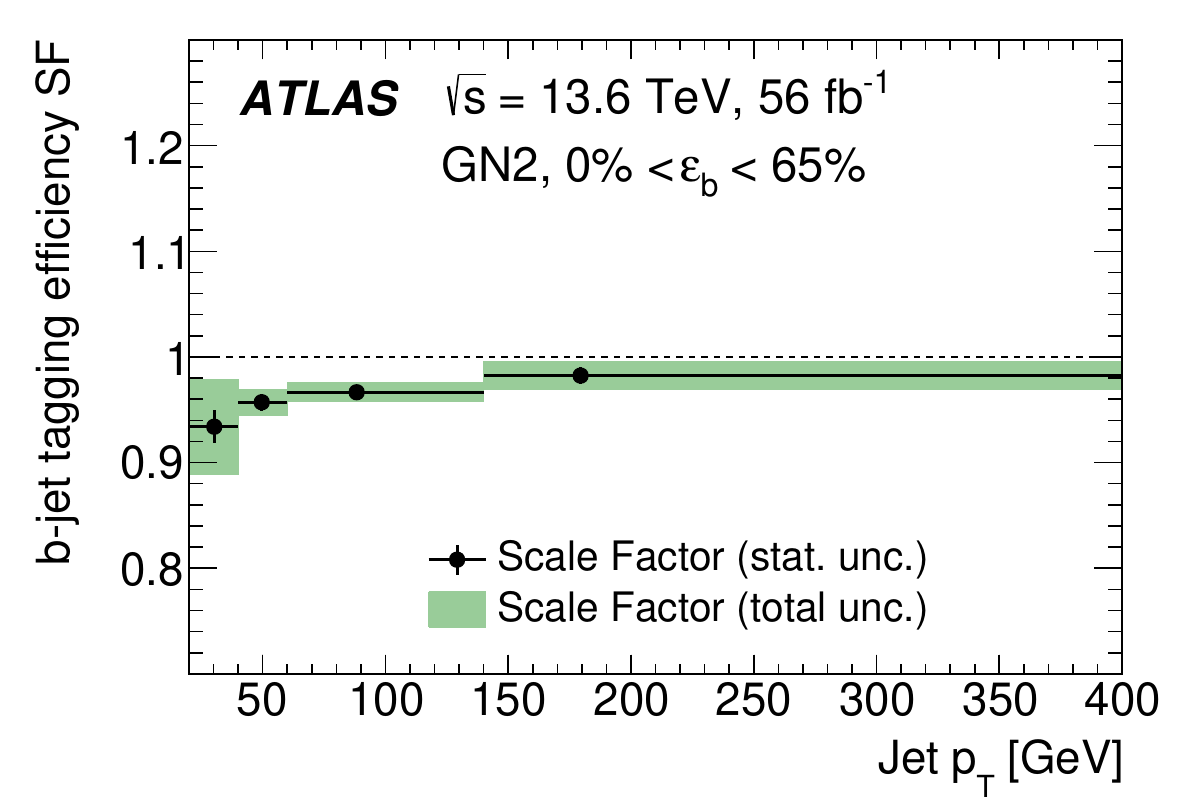}
}
\caption{
\bjet tagging efficiency simulation-to-data scale factors of GN2 as a function of jet \pt for all PCBT bins: (a) [100\%, 90\%], (b) [90\%, 85\%], (c) [85\%, 77\%], (d) [77\%, 70\%], (e) [70\%, 65\%] and (f) [65\%, 0\%]. Vertical error bars include data statistical uncertainties only while the green bands correspond to the sum in quadrature of data statistical, MC statistical and systematic uncertainties. The markers are located at the mean of the \bjet \pt distribution in each \pt bin.
}
\label{fig:result_SF_all}
\end{figure}

%
%
%
%

%
%
%
%

%

\begin{table}[htb!]
\setlength{\tabcolsep}{12pt}
\caption{Post-fit values of jet flavour composition correction factors $c_{bb}$, $c_{bl}$, $c_{lb}$ and $c_{ll}$ in intervals of leading and subleading jet \pt. Only data statistical uncertainties are quoted.}
\begin{adjustbox}{width=\textwidth, center}%
\begin{tabular}{llcccc}
\hline
Leading jet \pt[\GeV] & Subleading jet \pt[\GeV] & \multicolumn{4}{c}{Correction factors}\\
\cmidrule(lr){3-6}
&                      &  $c_{bb}$       & $c_{bl}$        & $c_{lb}$        & $c_{ll}$ \\
\toprule
$20<p_\text{T,1}<40$ & $20<p_\text{T,2}<40$ & 0.89 $\pm$ 0.03 & 0.88 $\pm$ 0.02 & 0.83 $\pm$ 0.03 & 1.02 $\pm$ 0.04 \\
\hline
$40<p_\text{T,1}<60$ & $20<p_\text{T,2}<40$ & 0.90 $\pm$ 0.02 & 0.85 $\pm$ 0.03 & 0.78 $\pm$ 0.05 & 0.97 $\pm$ 0.04 \\
& $40<p_\text{T,2}<60$ & 0.93 $\pm$ 0.02 & 0.89 $\pm$ 0.05 & 0.75 $\pm$ 0.06 & 1.18 $\pm$ 0.10 \\
\hline
$60<p_\text{T,1}<140$ & $20<p_\text{T,2}<40$ & 0.89 $\pm$ 0.02 & 0.85 $\pm$ 0.02 & 0.88 $\pm$ 0.05 & 0.98 $\pm$ 0.04 \\
& $40<p_\text{T,2}<60$ & 0.90 $\pm$ 0.01 & 0.81 $\pm$ 0.04 & 0.87 $\pm$ 0.04 & 1.08 $\pm$ 0.07 \\
& $60<p_\text{T,2}<140$ & 0.90 $\pm$ 0.01 & 0.83 $\pm$ 0.04 & 0.87 $\pm$ 0.04 & 1.04 $\pm$ 0.11 \\
\hline
$140<p_\text{T,1}<400$ & $20<p_\text{T,2}<40$ & 0.82 $\pm$ 0.03 & 0.82 $\pm$ 0.03 & 0.99 $\pm$ 0.10 & 1.18 $\pm$ 0.08 \\
& $40<p_\text{T,2}<60$ & 0.85 $\pm$ 0.03 & 0.66 $\pm$ 0.06 & 0.89 $\pm$ 0.09 & 1.23 $\pm$ 0.16 \\
& $60<p_\text{T,2}<140$ & 0.82 $\pm$ 0.02 & 0.80 $\pm$ 0.06 & 1.07 $\pm$ 0.07 & 1.30 $\pm$ 0.16 \\
& $140<p_\text{T,2}<400$ & 0.78 $\pm$ 0.03 & 1.21 $\pm$ 0.02 & 1.21 $\pm$ 0.13 & 1.06 $\pm$ 0.03\\
\hline
\end{tabular}
\end{adjustbox}
\label{tab:flavour_component_correction}
\end{table}


%
The uncertainty in the measurement of \bjet tagging efficiency for the [65\%, 0\%] PCBT bin is summarised in Figure \ref{fig:result_unc_65-0}.
The leading uncertainties in the low jet \pt region are the detector systematic uncertainties, which are dominated by the uncertainties in the energy scale and resolution of jets.
For $60<\textrm{jet}~\pt<140~\GeV$ where the number of events is the largest, the uncertainty breakdown is summarised for all PCBT bins in Table~\ref{tab:Breakdown_Table_yr_22_23_pt_60-140}. The total uncertainty in this interval of jet \pt ranges between 0.9\% and 7.8\%, depending on the PCBT bin.
For jet \pt larger than 140~\GeV, the data statistical uncertainty increases.

\begin{figure}[!htbp]
\centering
\includegraphics[width=0.6\textwidth]{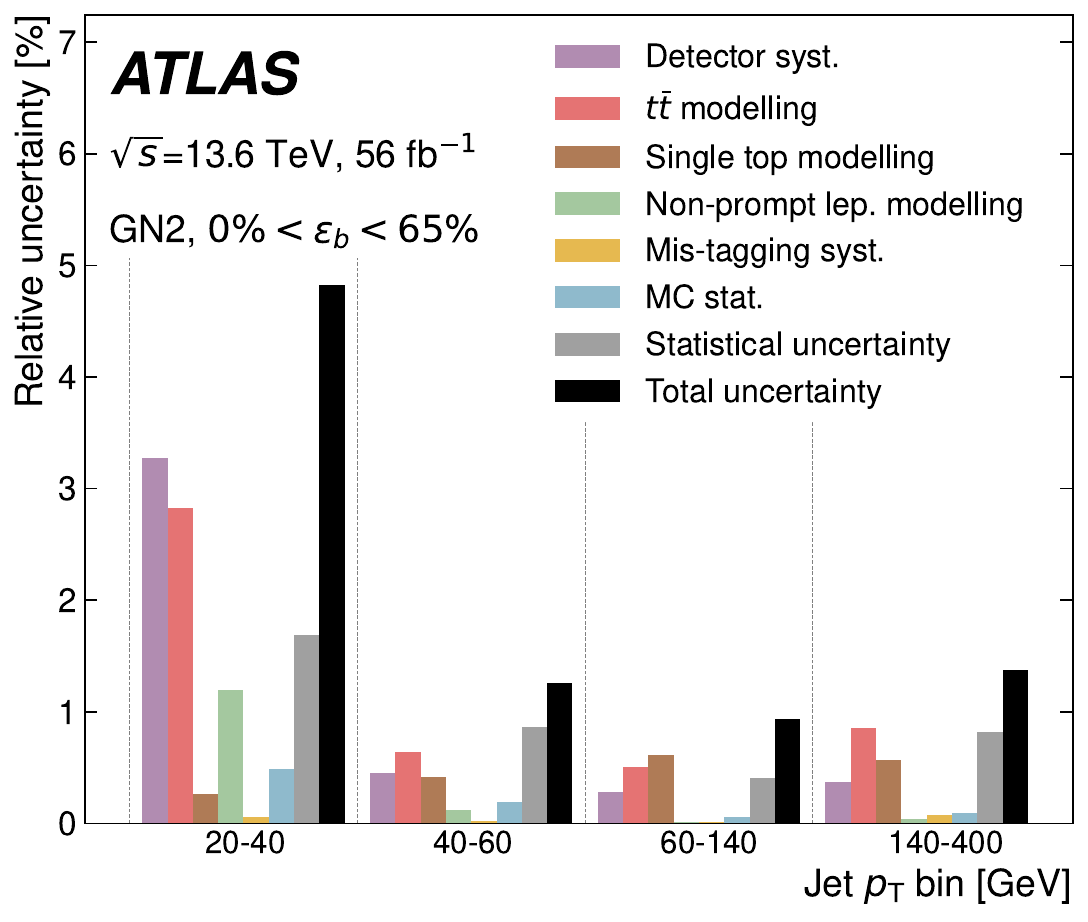}
\caption{Breakdown of the relative uncertainties impacting the \bjet tagging efficiency in the [65\%, 0\%] PCBT bin as a function of the jet \pt. Detector systematic uncertainties, \ttbar modelling uncertainties, single top modelling uncertainties, non-prompt lepton correction uncertainties, light-flavour jet mis-tagging uncertainties, MC statistical uncertainties, data statistical uncertainties, and total uncertainties are shown from left to right in each bin of jet \pt.}
\label{fig:result_unc_65-0}
\end{figure}

%
%

%

\begin{table}[htb!]
\centering
\caption{Relative uncertainty breakdown for the \bjet tagging efficiency in the $60<\pt<140~\GeV$ jet \pt bin. The PCBT bins are intervals of cumulative \bjet tagging efficiency.}
\begin{adjustbox}{width=\textwidth, center}
\begin{tabular}{lcccccc}
\hline
Source of uncertainty & \multicolumn{6}{c}{\makecell{Relative uncertainty [\%] on \\ \bjet tagging efficiency per PCBT bin}}\\
\cmidrule(lr){2-7}
&  [100\%, 90\%]  &  [90\%, 85\%]  &  [85\%, 77\%]  &  [77\%, 70\%]  &  [70\%, 65\%]  &  [65\%, 0\%] \\
\toprule
Detector syst.  & 2.2  & 0.3  & 0.2  & 0.3  & 0.2  & 0.3 \\
\ttbar modelling  & 4.5  & 0.7  & 0.5  & 0.5  & 0.5  & 0.5 \\
Single top modelling  & 5.2  & 0.1  & 0.5  & 0.6  & 0.5  & 0.6 \\
Non-prompt lep. modelling  & 0.1  & 0.1  & 0.2  & <0.1  & 0.1  & <0.1 \\
Mis-tagging syst.  & 0.6  & 0.5  & 0.2  & 0.2  & 0.1  & <0.1 \\
MC stat.  & 0.5  & 0.1  & 0.1  & 0.1  & 0.1  & 0.1 \\
Total systematic uncertainties  & 7.3  & 0.9  & 0.8  & 0.8  & 0.8  & 0.8 \\
Statistical uncertainty  & 3.0  & 1.9  & 1.5  & 1.6  & 1.8  & 0.4 \\
\toprule
\textbf{{Total uncertainty}}  & 7.8  & 2.1  & 1.7  & 1.7  & 1.9  & 0.9 \\
\hline
\end{tabular}
\end{adjustbox}
\label{tab:Breakdown_Table_yr_22_23_pt_60-140}
\end{table}



%
%
\FloatBarrier

\section{Conclusion}
\label{sec:conclusion}
%

The \bjet performance of the GN2 algorithm is studied with data recorded by the ATLAS experiment during the years 2022 and 2023 of \RunThr at a centre-of-mass energy of 13.6~\TeV, corresponding to an integrated luminosity of 56~\ifb.

The measurement of the \bjet tagging efficiency of the GN2 algorithm in \pp collisions relies on a sample enriched in dileptonic \ttbar events, which is obtained by selecting events with exactly two jets, a single electron and a single muon of opposite electric charge. These selected events are further categorised according to the \pt of the two jets, the invariant mass of the two jet--lepton pairs, and the PCBT bins of the two jets. The categorised events are used as input to a likelihood fit that simultaneously infers the \bjet tagging efficiencies and the jet flavour composition of the data. The \bjet tagging efficiencies are measured for jets with a transverse momentum between 20 and 400~\GeV, providing \bjet tagging efficiencies in each PCBT bin. Simulation-to-data scale factors are then computed in PCBT bins as a function of jet \pt by dividing the measured efficiency in collision data by the expected value in the simulation. The simulation-to-data scale factors obtained from the measurements are between 0.9 and 1.3. The measured \bjet tagging efficiencies have a total uncertainty of around 1\% for jets with transverse momentum larger than 60~\GeV in the [65\%, 0\%] interval of cumulative efficiency. The precision of the measurement at low jet \pt is mostly limited by the uncertainties in jet energy calibration and physics modelling, while at high jet \pt the total uncertainty receives a larger contribution from the statistical uncertainty of the data. The jet flavour composition correction factors inferred from the data events are also reported for all intervals of jet \pt.

These results validate the outstanding \bjet tagging performance of the GN2 algorithm used by the ATLAS experiment during the \RunThr data-taking. The central values of the \bjet tagging efficiency scale factors of the GN2 algorithm are provided to ATLAS physics analyses, along with the uncertainties and correlation scheme of uncertainties sources across jet \pt and PCBT bins.
%


%
%
%
%
%
\section*{Acknowledgements}
%

%

%
%

%
%

We thank CERN for the very successful operation of the LHC and its injectors, as well as the support staff at
CERN and at our institutions worldwide without whom ATLAS could not be operated efficiently.

The crucial computing support from all WLCG partners is acknowledged gratefully, in particular from CERN, the ATLAS Tier-1 facilities at TRIUMF/SFU (Canada), NDGF (Denmark, Norway, Sweden), CC-IN2P3 (France), KIT/GridKA (Germany), INFN-CNAF (Italy), NL-T1 (Netherlands), PIC (Spain), RAL (UK) and BNL (USA), the Tier-2 facilities worldwide and large non-WLCG resource providers. Major contributors of computing resources are listed in Ref.~\cite{ATL-SOFT-PUB-2026-001}.

We gratefully acknowledge the support of ANPCyT, Argentina; YerPhI, Armenia; ARC, Australia; BMWFW and FWF, Austria; ANAS, Azerbaijan; CNPq and FAPESP, Brazil; NSERC, NRC and CFI, Canada; CERN; ANID, Chile; CAS, MOST, NSFC and FRFCU, China; Minciencias, Colombia; MEYS CR, Czech Republic; DNRF and DNSRC, Denmark; IN2P3-CNRS and CEA-DRF/IRFU, France; SRNSFG, Georgia; BMFTR, HGF and MPG, Germany; GSRI, Greece; RGC and Hong Kong SAR, China; ICHEP and Academy of Sciences and Humanities, Israel; INFN, Italy; MEXT and JSPS, Japan; CNRST, Morocco; NWO, Netherlands; RCN, Norway; MNiSW, Poland; FCT, Portugal; MNE/IFA, Romania; MSTDI, Serbia; MSSR, Slovakia; ARIS and MVZI, Slovenia; DSI/NRF, South Africa; MICIU/AEI, Spain; SRC and Wallenberg Foundation, Sweden; SERI, SNSF and Cantons of Bern and Geneva, Switzerland; NSTC, Taipei; TENMAK, T\"urkiye; STFC/UKRI, United Kingdom; DOE and NSF, United States of America.

Individual groups and members have received support from BCKDF, CANARIE, CRC and DRAC, Canada; CERN-CZ, FORTE and PRIMUS, Czech Republic; COST, ERC, ERDF, Horizon 2020 and Marie Sk{\l}odowska-Curie Actions, European Union; Investissements d'Avenir Labex, Investissements d'Avenir Idex and ANR, France; DFG and AvH Foundation, Germany; Herakleitos, Thales and Aristeia programmes co-financed by EU-ESF and the Greek NSRF, Greece; BSF-NSF and MINERVA, Israel; NCN and NAWA, Poland; La Caixa Banking Foundation, CERCA and AGAUR programs from Generalitat de Catalunya and PROMETEO and GenT Programmes Generalitat Valenciana, Spain; G\"{o}ran Gustafssons Stiftelse, Sweden; The Royal Society and Leverhulme Trust, United Kingdom; Eric and Wendy Schmidt Fund for Strategic Innovation, United States of America.

In addition, individual members wish to acknowledge support from Chile: Agencia Nacional de Investigaci\'on y Desarrollo (ANID FONDECYT reg. 1230987, FONDECYT 1230812, FONDECYT 1240864, Fondecyt 3240661, Fondecyt Regular 1240721); China: Fundamental Research Funds for the Central Universities (010-63263105), Chinese Ministry of Science and Technology (MOST-2023YFA1605700, MOST-2023YFA1609300), National Natural Science Foundation of China (NSFC 12275265, NSFC-W2543005); Czech Republic: Czech Science Foundation (GACR - 24-11373S), Ministry of Education Youth and Sports (ERC-CZ-LL2327, FORTE CZ.02.01.01/00/22\_008/0004632), PRIMUS Research Programme (PRIMUS/21/SCI/017); EU: H2020 European Research Council (ERC - 101002463); European Union: European Research Council (BARD No. 101116429, ERC - 101219398, ERC - 948254, ERC 101089007), European Regional Development Fund (HE COFUND GA No.101081355, ERDF), Marie Sklodowska-Curie Actions (GAP-101168829); France: Agence Nationale de la Recherche (ANR-21-CE31-0013, ANR-22-EDIR-0002, ANR-24-CE31-0504-01); Germany: Deutsche Forschungsgemeinschaft (DFG - 469666862); China: Research Grants Council (GRF); Italy: Istituto Nazionale di Fisica Nucleare (LHC-MIUR - 28003/2025), Ministero dell'Università e della Ricerca (NextGenEU 153D23001490006 M4C2.1.1, NextGenEU I53D23000820006 M4C2.1.1, SOE2024\_0000023); Japan: Japan Society for the Promotion of Science (JSPS KAKENHI  JP25H0063, JSPS KAKENHI JP22H01227, JSPS KAKENHI JP22H04944, JSPS KAKENHI JP22KK0227, JSPS KAKENHI JP24K23939, JSPS KAKENHI JP24KK0251, JSPS KAKENHI JP25H00650, JSPS KAKENHI JP25H01291, JSPS KAKENHI JP25K01011, JSPS KAKENHI JP25K01023); Poland: Polish National Science Centre (NCN 2021/42/E/ST2/00350, NCN OPUS 2023/51/B/ST2/02507, NCN OPUS nr 2022/47/B/ST2/03059, NCN UMO-2019/34/E/ST2/00393, UMO-2022/47/O/ST2/00148, UMO-2023/49/B/ST2/04085, UMO-2023/51/B/ST2/00920, UMO-2024/53/N/ST2/00869); Spain: Agència de Gestió d'Ajuts Universitaris i de Recerca. (AGAUR - 2023 BP 00141), Ministry of Science and Innovation (RYC2019-028510-I, RYC2020-030254-I, RYC2021-031273-I, RYC2022-038164-I), Ministerio de Ciencia, Innovación y Universidades/Agencia Estatal de Investigaci\'on (EU NextGenerationEU (PRTR-C17.I1), PID2022-142604OB-C22); Sweden: Carl Trygger Foundation (Carl Trygger Foundation CTS 22:2312), Swedish Research Council (Swedish Research Council 2023-04654, VR 2021-03651, VR 2022-03845, VR 2022-04683, VR 2023-03403, VR 2024-05451, VR 2025-05940), Knut and Alice Wallenberg Foundation (KAW 2023.0366); Switzerland: Swiss National Science Foundation (SNSF - PCEFP2\_194658); United Kingdom: The Binks Trust, Royal Society (NIF-R1-231091); United States of America: U.S. Department of Energy (ECA DE-AC02-76SF00515), John Templeton Foundation (John Templeton Foundation 63206), Neubauer Family Foundation.

%
%


%
%
%
%
%
%

%
%
%
\clearpage
\printbibliography
\clearpage
 
\begin{flushleft}
\hypersetup{urlcolor=black}
{\Large The ATLAS Collaboration}

\bigskip

\AtlasOrcid[0000-0002-6665-4934]{G.~Aad}$^\textrm{\scriptsize 102}$,
\AtlasOrcid[0000-0001-7616-1554]{E.~Aakvaag}$^\textrm{\scriptsize 17}$,
\AtlasOrcid[0000-0002-5888-2734]{B.~Abbott}$^\textrm{\scriptsize 121}$,
\AtlasOrcid[0000-0002-0287-5869]{S.~Abdelhameed}$^\textrm{\scriptsize 83b}$,
\AtlasOrcid[0000-0002-1002-1652]{K.~Abeling}$^\textrm{\scriptsize 54}$,
\AtlasOrcid[0000-0001-5763-2760]{N.J.~Abicht}$^\textrm{\scriptsize 48}$,
\AtlasOrcid[0000-0002-8496-9294]{S.H.~Abidi}$^\textrm{\scriptsize 30}$,
\AtlasOrcid[0009-0003-6578-220X]{M.~Aboelela}$^\textrm{\scriptsize 44}$,
\AtlasOrcid[0000-0002-9987-2292]{A.~Aboulhorma}$^\textrm{\scriptsize 36e}$,
\AtlasOrcid[0000-0001-5329-6640]{H.~Abramowicz}$^\textrm{\scriptsize 154}$
% ...
\bigskip
\\

% ...
$^{17}$Department for Physics and Technology, University of Bergen, Bergen; Norway.\\
% ...
$^{30}$Physics Department, Brookhaven National Laboratory, Upton NY; United States of America.\\
% ...
$^{36}$$^{(a)}$Facult\'e des Sciences Ain Chock, Universit\'e Hassan II de Casablanca;$^{(b)}$Facult\'{e} des Sciences, Universit\'{e} Ibn-Tofail, K\'{e}nitra;$^{(c)}$Facult\'e des Sciences Semlalia, Universit\'e Cadi Ayyad, LPHEA-Marrakech;$^{(d)}$LPMR, Facult\'e des Sciences, Universit\'e Mohamed Premier, Oujda;$^{(e)}$Facult\'e des sciences, Universit\'e Mohammed V, Rabat;$^{(f)}$Institute of Applied Physics, Mohammed VI Polytechnic University, Ben Guerir; Morocco.\\
% ...
$^{44}$Physics Department, Southern Methodist University, Dallas TX; United States of America.\\
% ...
$^{48}$Fakult\"{a}t Physik, Technische Universit{\"a}t Dortmund, Dortmund; Germany.\\
% ...
$^{54}$II. Physikalisches Institut, Georg-August-Universit\"{a}t G\"ottingen, G\"ottingen; Germany.\\
% ...
$^{83}$$^{(a)}$Khalifa University of Science and Technology, Abu Dhabi;$^{(b)}$New York University Abu Dhabi, Abu Dhabi;$^{(c)}$United Arab Emirates University, Al Ain;$^{(d)}$University of Sharjah, Sharjah; United Arab Emirates.\\
% ...
$^{102}$CPPM, Aix-Marseille Universit\'e, CNRS/IN2P3, Marseille; France.\\
% ...
$^{121}$Homer L. Dodge Department of Physics and Astronomy, University of Oklahoma, Norman OK; United States of America.\\
% ...
$^{154}$Raymond and Beverly Sackler School of Physics and Astronomy, Tel Aviv University, Tel Aviv; Israel.\\
% ...
$^{a}$ Also at Affiliated with an institute formerly covered by a cooperation agreement with CERN.\\
$^{b}$ Also at An-Najah National University, Nablus; Palestine.\\
$^{c}$ Also at Borough of Manhattan Community College, City University of New York, New York NY; United States of America.\\
$^{d}$ Also at Center for Interdisciplinary Research and Innovation (CIRI-AUTH), Thessaloniki; Greece.\\
$^{e}$ Also at Centre of Physics of the Universities of Minho and Porto (CF-UM-UP); Portugal.\\
$^{f}$ Also at CERN, Geneva; Switzerland.\\
% ...

\end{flushleft}


%

%
%

%

%
%
%
%

\end{document}